\documentclass{aa}
\usepackage[varg]{txfonts}
\usepackage{graphicx}

\bibpunct{(}{)}{;}{a}{}{,} 

\begin{document}

\title{Discovery of a directly imaged disk in scattered light around the Sco-Cen member \object{Wray~15-788}\thanks{Based on observations collected at the European Organisation for Astronomical Research in the Southern Hemisphere under ESO programs 099.C-0698(A), 0101.C-0153(A), and 0101.C-0464(A).}}

\author{
A.~J.~Bohn\inst{1}
\and M.~A.~Kenworthy\inst{1}
\and C.~Ginski\inst{1,2}
\and M.~Benisty\inst{3,4}
\and J.~de~Boer\inst{1}
\and C.~U.~Keller\inst{1}
\and E.~E.~Mamajek\inst{5,6}
\and T.~Meshkat\inst{7}
\and G.~A.~Muro-Arena\inst{2}
\and M.~J.~Pecaut\inst{8}
\and F.~Snik\inst{1}
\and S.~G.~Wolff\inst{1}
\and M.~Reggiani\inst{9,10}
}

\institute{Leiden Observatory, Leiden University, PO Box 9513, 2300 RA Leiden, The Netherlands\\
              \email{bohn@strw.leidenuniv.nl}
              \and Sterrenkundig Instituut Anton Pannekoek, Science Park 904, 1098 XH Amsterdam, The Netherlands
       \and Unidad Mixta Internacional Franco-Chilena de Astronom\'{i}a (CNRS, UMI 3386), Departamento de Astronom\'{i}a, Universidad de Chile, Camino El Observatorio 1515, Las Condes, Santiago, Chile
       \and Univ. Grenoble Alpes, CNRS, IPAG, 38000 Grenoble, France.
       \and Jet Propulsion Laboratory, California Institute of Technology, 4800 Oak Grove Drive, M/S 321-100, Pasadena, CA, 91109, USA
       \and Department of Physics \& Astronomy, University of Rochester, Rochester, NY 14627, USA
       \and IPAC, California Institute of Technology, M/C 100-22, 1200 East California Boulevard, Pasadena, CA 91125, USA
       \and Rockhurst University, Department of Physics, 1100 Rockhurst Road, Kansas City, MO 64110, USA
       \and Space sciences, Technologies and Astrophysics Research (STAR) Institute, Université de Liège, Allée du 6 Août 17, Bat. B5C, 4000 Liège, Belgium
       \and Institute of Astronomy, KU Leuven, Celestijnenlaan 200D, B-3001 Leuven, Belgium
       }

\date{Received October 28, 2018 / Accepted February 20, 2019}

\abstract 
{
Protoplanetary disks are the birth environments of planetary systems. 
Therefore, the study of young, circumstellar environments is essential to understanding the processes taking place in planet formation and the evolution of planetary systems.
} 
{
We detect and characterize circumstellar disks and potential companions around solar-type, pre-main sequence stars in the Scorpius-Centaurus association (Sco-Cen).
} 
{
As part of our ongoing survey we carried out high-contrast imaging with VLT/SPHERE/IRDIS to obtain polarized and total intensity images of the young ($11^{+16}_{-7}$\,Myr old) K3IV star \object{Wray~15-788} within the Lower Centaurus Crux subgroup of Sco-Cen. 
For the total intensity images, we remove the stellar halo via an approach based on reference star differential imaging in combination with principal component analysis.
} 
{
Both total intensity and polarimetric data resolve a disk around the young, solar-like Sco-Cen member \object{Wray~15-788}.
Modeling of the stellar spectral energy distribution suggests that this is a protoplanetary disk at a transition stage.
We detect a bright outer ring at a projected separation of $\sim$370\,mas ($\approx$\,56\,au), hints of inner substructures at $\sim$170\,mas ($\approx$\,28\,au), and a gap in between.
Within a position angle range of only 60\degr\,<\,$\phi$\,<\,240\degr, we are confident at the 5$\sigma$ level that we detect actual scattered light flux from the outer ring of the disk; the remaining part is indistinguishable from background noise.
For the detected part of the outer ring we determine a disk inclination of $i$\,=\,21\degr$\,\pm\,$6\degr and a position angle of $\varphi$\,=\,76\degr$\,\pm\,$16\degr.
Furthermore, we find that \object{Wray~15-788} is part of a binary system with the A2V star \object{HD~98363} at a separation of $\sim$50\arcsec ($\approx$\,6900\,au).
} 
{
The detection of only half of the outer ring might be due to shadowing by a misaligned inner disk. 
A potential substellar companion can cause the misalignment of the inner structures and can be responsible for clearing the detected gap from scattering material.
However, we cannot rule out the possibility of a non-detection due to our limited signal-to-noise ratio, combined with brightness azimuthal asymmetry.
From our data we can exclude companions more massive than 10\,$M_\text{jup}$ within the gap at a separation of $\sim$230\,mas ($\approx$\,35\,au).
Additional data are required to characterize the disk's peculiar morphology and to set tighter constraints on the potential perturber's orbital parameters and mass.
}


\maketitle

\section{Introduction}
\label{sec:introduction}

In the past few years, the second generation of high-contrast imaging instruments such as the Spectro-Polarimetric High-contrast Exoplanet REsearch \citep[SPHERE,][]{Beuzit2019} instrument and the Gemini Planet Imager \citep[GPI,][]{GPI} have resolved and characterized several disks around young, pre-main sequence stars \citep[e.g.,][]{Avenhaus2018,Millar-Blanchaer2017}.
These range from warm, gas-rich protoplanetary disks around young stars of ages usually lower than 10\,Myr \citep{Andrews2012} to cold debris disks around more evolved stars where the primordial gas has already dissipated \citep{Matsuyama2003,Wyatt2003}.
Since planets form within protoplanetary disks \citep{Goldreich1973}, the characterization of circumstellar environments and the search for planetary mass companions is closely related. 
The study of young stellar systems, therefore, gives us an understanding of the initial conditions of planet formation.

With a mean distance of $\sim$130\,pc \citep{deZeeuw1999} and an average age of 14\,$\pm$3\,Myr \citep{Pecaut2016}, the Scorpius-Centaurus association \citep[Sco-Cen,][]{deZeeuw1999} is one of the closest sites of recent star formation to the Sun.
Therefore, Sco-Cen is an ideal region when it comes to the search for young, luminous planets or protoplanetary and early debris disks.
\citet{Pecaut2016} identified and characterized 156 new K-type star members of Sco-Cen.
One object in this sample is the emission-line star \object{Wray~15-788} (\object{2MASS~J11175186-6402056}, \object{Hen~3-632}), which is located in the Lower Centaurus-Crux (LCC) subgroup of Sco-Cen \citep{Mamajek2013,Pecaut2016}.
It was discovered as an H$\alpha$ emission object by \cite{Wray1966} and was confirmed within the study of southern emission line stars by \citet{Henzie1976}.
The star is of spectral type K3IVe, has a mass of 1.2\,$M_\sun$ \citep{Pecaut2016}, and a distance of $139.7\pm0.5$\,pc \citep{GAIA2018}. 
In addition, \citet{Pecaut2016} determined an age of 4\,Myr, which is likely an underestimate.
A more accurate age may be obtained by using evolutionary models that include magnetic fields \citep{Feiden16} as presented in Sect.~\ref{subsec:binary} of this work.
Table~\ref{tbl:WRAY15788_parameters} summarizes the most important stellar parameters of \object{Wray~15-788}.
\begin{table*}
\caption{Stellar parameters of \object{Wray~15-788} and \object{HD~98363}.}
\label{tbl:WRAY15788_parameters}
\def\arraystretch{1.2}
\setlength{\tabcolsep}{12pt}
\centering
\begin{tabular}{@{}llll@{}}
\hline\hline
Parameter & \multicolumn{2}{c}{Value} & Reference(s)\\ 
 & \object{Wray~15-788} & \object{HD~98363} & \\
\hline
Right Ascension (J2000) & 11:17:51.87 & 11:17:58.14 & (1) \\
Declination (J2000) & -64:02:05.60 & -64:02:33.35 & (1) \\
Spectral Type & K3IVe & A2V & (2,3) \\
Mass [$M_\sun$] & $1.26^{+0.07}_{-0.22}$ & $1.92^{+0.08}_{-0.08} $ & (2,4) \\
Effective Temperature [K] & $4\,549^{+225}_{-215}$ & $8\,830^{+331}_{-319}$ & (2,4)\\
Luminosity [$L_\sun$] & $0.91^{+0.07}_{-0.06}$ & $14.96^{+1.44}_{-1.32}$ & (2,4)\\
Age\tablefootmark{a} [Myr] & 11$^{+16}_{-7}$ & $11^{+16}_{-7}$ & (2,4) \\
Parallax [mas] & $7.159\pm0.027$ & $7.215\pm 0.034$ & (1) \\
Distance [pc] & $139.126\pm 0.52$ & $138.044\pm 0.66$ & (1,5) \\
Proper motion (RA) [mas] & $-28.583\pm0.042$ & $-28.491\pm0.053$ & (1) \\
Proper motion (Dec) [mas] & $-1.411\pm0.040$ & $-0.795\pm0.051$ & (1) \\
$V$ [mag]& $11.89\pm0.08$ & $7.85\pm0.01$ & (2,6,7) \\
$B-V$ [mag] & $1.11\pm0.09$ & $0.18\pm0.01$ & (2,6,7) \\
$J$ [mag] & $9.39\pm0.02$ & $7.53\pm0.02$& (2,8) \\
$H$ [mag] & $8.59\pm0.04$ & $ 7.48\pm0.03$ & (2,8) \\
$K_\text{s}$ [mag] & $8.18\pm0.03$ & $7.50\pm0.02$ & (2,8) \\
$W1$ [mag] & $7.75\pm0.02$ & $7.36\pm0.03$ & (2,9) \\
$W2$ [mag] & $7.49\pm0.02$ & $7.43\pm0.02$ & (2,9) \\
$W3$ [mag] & $6.42\pm0.02$ & $6.93\pm0.02$ & (2,9) \\
$W4$ [mag] & $3.88\pm0.02$ & $4.64\pm0.02$ & (2,9) \\
\hline
\end{tabular}
\tablefoot{
\tablefoottext{a}{
The primary, \object{HD~98363}, has a most likely age of 11 Myr, but since the error bars of the primary overlap the main sequence in the Hertzsprung-Russell diagram, the 95\% confidence range of 22 Myr to 480 Myr does not contain the mode.
Given its membership in Sco-Cen, this is not a useful age constraint. 
The secondary, Wray 15-788, is in a stage of evolution where we can place meaningful limits on the age, so we adopt the system age as that of the secondary, 11 Myr with 95\% CL range of $11^{+16}_{-7}\,Myr$.
}
}
\tablebib{
(1)~\citet{GAIA2018}; (2)~\citet{Pecaut2016}; (3)~\citet{Houk75}; (4)~Sect.~\ref{subsec:binary} of this work; (5)~\citet{BailerJones18}; (6)~\citet{Henden2012}; (7)~\citet{Hog2000}; (8)~\citet{Cutri2012a}; (9)~\citet{Cutri2012b}
}
\end{table*}

In Sect.~\ref{sec:observations} we describe the SPHERE data we obtained on \object{Wray~15-788} and Sect.~\ref{sec:data_reduction} explains our applied data reduction techniques. 
Thereafter, we present our observational results in Sect.~\ref{sec:results} and an analysis of these data is given in Sect.~\ref{sec:analysis}.
Furthermore, we show the association of \object{Wray~15-788} as a comoving companion to the main sequence star \object{HD 98363} and derive new estimates for ages and masses of both objects.
A model of the stellar spectral energy distribution (SED) is also presented in Sect.~\ref{sec:analysis}.
Finally, we discuss our results in Sect.~\ref{sec:discussion} and present the conclusions of the article in Sect.~\ref{sec:conclusions}.

\section{Observations}
\label{sec:observations}

All our observations were performed with SPHERE, which is mounted on the Naysmith platform of Unit 3 telescope (UT3) at ESO's VLT.
To obtain diffraction limited data, SPHERE is assisted by the SAXO extreme adaptive optics system \citep{Fusco2006,Petit2014}.
In particular, we made use of the infrared dual-band imager and spectrograph \citep[IRDIS;][]{Dohlen2008}, which was operated in both dual-polarization imaging \citep[DPI;][]{Langlois2014} and classical imaging \citep[CI;][]{Vigan2010} modes to obtain high-contrast polarized and total intensity images of the system. 
A detailed description of the observations is presented in Table \ref{tbl:WRAY15788_observations}.
\begin{table*}
\caption{Observations of \object{Wray~15-788} carried out with SPHERE/IRDIS.}
\label{tbl:WRAY15788_observations}
\def\arraystretch{1.2}
\setlength{\tabcolsep}{12pt}
\centering
\begin{tabular}{@{}llllllll@{}}
\hline\hline
Observation date & Mode\tablefootmark{a} & Filter & NDIT$\times$DIT\tablefootmark{b} & $\Delta\pi$\tablefootmark{c} & $\langle\omega\rangle$\tablefootmark{d} & $\langle X\rangle$\tablefootmark{e} & $\langle\tau_0\rangle$\tablefootmark{f} \\ 
(yyyy-mm-dd) & & & (1$\times$s) & (\degr) & (\arcsec) & & (ms)\\
\hline
2018-05-14 & CI & $H$ & 4$\times$32 & 0.86 & 0.86 & 1.30 & 2.55 \\
2018-05-14 & CI & \textit{K}$_\textit{s}$ & 4$\times$32 & 0.87 & 0.85 & 1.30 & 2.15 \\
2018-06-05 & DPI & $H$ & 4$\times$64 & - & 0.99 & 1.30 & 1.48 \\
\hline
\end{tabular}
\tablefoot{
\tablefoottext{a}{Observation mode is either classical imaging (CI) or dual-polarization imaging (DPI).}
\tablefoottext{b}{NDIT describes the number of dithering positions and DIT is the detector integration time per dithering position.}
\tablefoottext{c}{$\Delta\pi$ describes the amount of field rotation during the observation, if it is carried out in pupil-stabilized mode (only valid for CI observations).}
\tablefoottext{d}{$\langle\omega\rangle$ denotes the average seeing conditions during the observation.}
\tablefoottext{e}{$\langle X\rangle$ denotes the average airmass during the observation.}
\tablefoottext{f}{$\langle\tau_0\rangle$ denotes the average coherence time during the observation.}
}
\end{table*}

\subsection{Classical imaging}

The CI observations (PI: M.~A.~Kenworthy) were obtained on May~14, 2018, within a larger program looking for planetary mass companions around solar-type stars in Sco-Cen (Bohn et al. in prep).
The target was observed in good weather conditions with two broadband filters in the $H$ and \textit{K}$_\textit{s}$ band (Filter IDs: \texttt{BB\_H}, \texttt{BB\_Ks}) for 128\,s each.
The central wavelengths of the filters are $\lambda_\text{c}^{H}$=1625.5\,nm and $\lambda_\text{c}^{{K}_{s}}$=2181.3\,nm with bandwidths of $\Delta\lambda^{H}$=291.0\,nm and $\Delta\lambda^{{K}_{s}}$=313.5\,nm, respectively.
To reduce the effect of bad detector pixels, a dither pattern on a 2$\times$2 grid with 1pixel spacing was applied during the observation.
Additionally, an apodized pupil Lyot coronagraph \citep{Soummer2005,Carbillet2011,Guerri2011} with a diameter of 185\,mas (Coronagraph ID: \texttt{N\_ALC\_YJH\_S}) was used to block the central flux of the star.
The observations were carried out in pupil-stabilized mode, but the amount of field rotation during the observation was less than 1\degr.
To model the thermal sky and instrument background, an additional exposure with the science setup was taken at an offset sky position without any source.
Center frames were obtained, for which a sinusoidal pattern was applied to the deformable mirror in order to create four calibration spots around the target's position behind the coronagraphic mask.
In addition, we obtained unsaturated, non-coronagraphic flux frames of the star with a neutral density filter (Filter ID: \texttt{ND\_1.0}) in place to avoid saturation of the detector.

\subsection{Dual-polarization imaging}

The DPI observation (PI: M.~Benisty) was carried out on the night of June~5, 2018, under very poor weather conditions. 
We obtained one polarimetric cycle, which consists of one image for each of the four half-wave plate positions ($0\degr$, $45\degr$, $22.5\degr$, and $67.5\degr$) with an exposure time of 64\,s each.
Furthermore, we applied the same coronagraph and broadband filter in $H$ band as was used for the CI observations. 
In a similar manner as described before, we also obtained additional center and sky frames for the DPI observation. 
The DPI cycle was conducted in field-stabilized mode.

\section{Data reduction}
\label{sec:data_reduction}

Both CI and DPI data were reduced by a personal processing pipeline based on the new release of the PynPoint package \citep{Stolker2019}. 
This included basic image processing steps such as flat fielding and sky subtraction for both CI and DPI data. 
Furthermore, a simple bad pixel correction was applied by a 5$\sigma$ box filtering algorithm \citep[based on the IDL routine of][]{Varosi1993}.

\subsection{Classical imaging}
\label{subsec:data_reduction_CI}

The dithering offsets of the science images to the center frame were registered and all frames were aligned accordingly.
Afterwards, the aligned science images were centered with respect to the star's position behind the coronagraph.
This position was determined as the center of the four calibration spots within the additionally obtained center frame \citep[see][]{Langlois2013}.
Because IRDIS was operated in CI mode, we obtained two copies of the coronagraphic stellar point spread function (PSF) simultaneously for each exposure \citep[see][]{Dohlen2008}.
To compensate for bad pixel introduced noise, we averaged the two centered PSFs from both detector sides for each individual exposure.
Finally, we removed the stellar halo and instrumental artifacts by an approach based on reference star differential imaging \citep[RDI,][]{Smith1984,Lafreniere2007}. 
Within a larger survey for planets around solar-type stars (PI: M.~A.~Kenworthy), in Sco-Cen we observed 26 and 12 stars in $H$ and $K_s$ band, respectively.
A detailed list of these reference stars and the observing conditions is presented in Appendix \ref{sec:ref_star_library}.
The stars are very similar to \object{Wray~15-788} in terms of spectral type, mass, age, distance, position on sky, and apparent magnitude.
Furthermore, they were observed with exactly the same observational setup as for \object{Wray~15-788}. 
Therefore, we created a library from these reference targets, on which we applied principal component analysis \citep[PCA;][]{PynPoint,Soummer2012}. 
Thereafter, the PSF of \object{Wray~15-788} in each science frame was modeled as linear combination of the first $m$ principal components (PCs) from the reference library \citep[RDI+PCA; e.g.,][]{Choquet2014}.
These PSF-models were subtracted from the science images, the residuals were de-rotated according to their parallactic angle and median combined.
For characterization of disks at low inclination, this technique has proven superior to algorithms based on angular differential imaging \citep{Marois2006}, which leads to undesirable self-subtraction effects from radial symmetric parts of the disk \citep{Choquet2014}.
An additional constant rotation of 135\fdg99 in the counterclockwise direction was applied to correct for the instrument's offset angle included to align the pupil with the Lyot stop\footnote{This value is obtained from the latest version of the instrument manual: \url{https://www.eso.org/sci/facilities/paranal/instruments/sphere/doc.html}}.
We used the general astrometric solution for IRDIS with a plate scale of 12.251\,$\pm$\,0.009\,mas~per~pixel and 12.265\,$\pm$\,0.009\,mas~per~pixel for $H$ and \textit{K}$_\textit{s}$ band, respectively, as well as a true north correction of $-1\fdg75\pm0\fdg08$ according to \citet{Maire2016}.

\subsection{Dual-polarization imaging}
\label{subsec:data_reduction_DPI}

The reduction of the DPI data was carried out following the description given in \cite{Ginski2016}.

\begin{figure*}
\centering
\includegraphics[width=17cm]{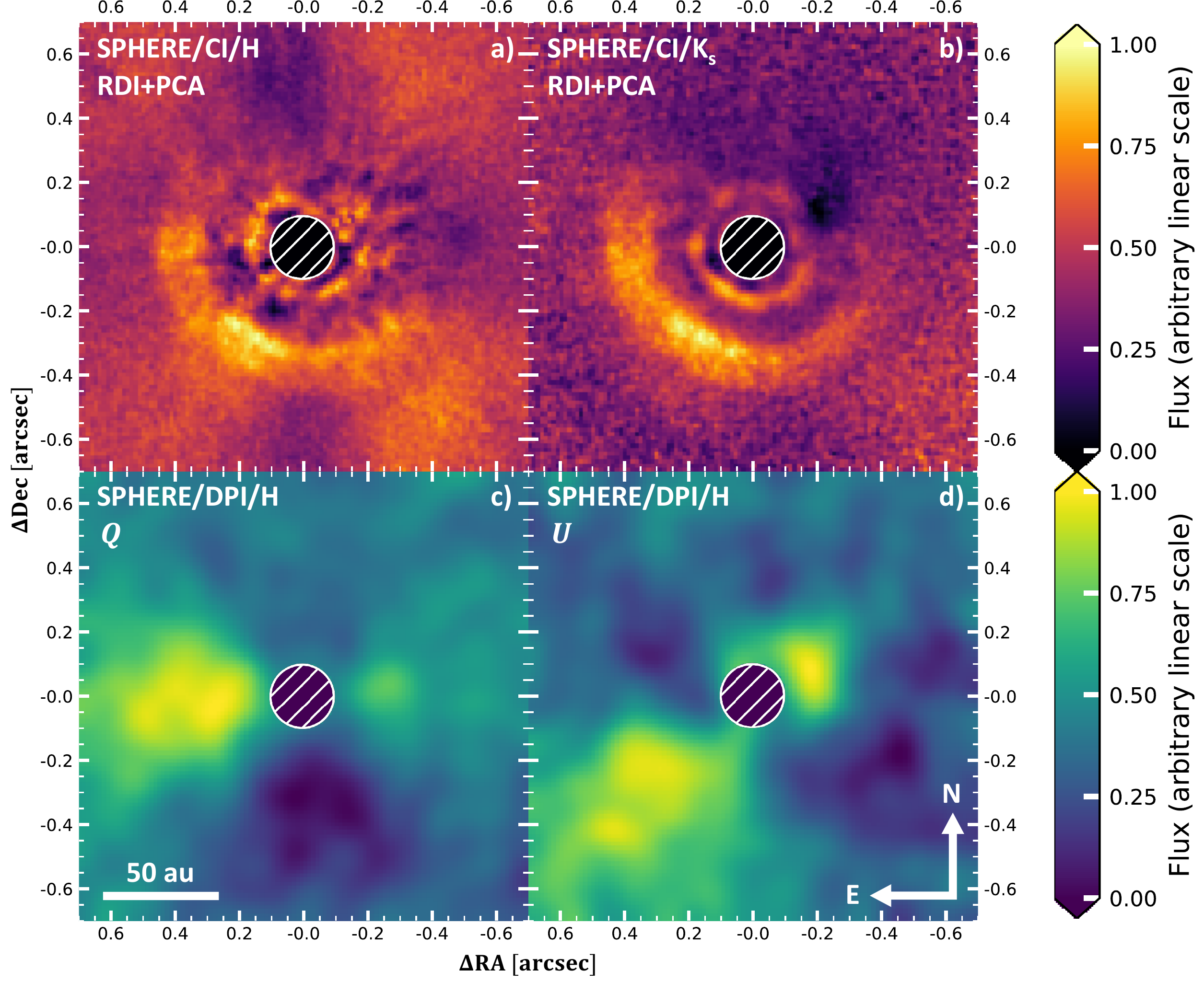}
\caption{
Reduced SPHERE images of \object{Wray~15-788}. 
All frames show the same region on the sky with a field of view of approximately 1\farcs39$\times$1\farcs39.
The star is positioned in the center of each image. 
An artificial mask with a diameter of $\sim$196\,mas is applied to obscure the coronagraph and leaking flux close to it.
The images are scaled with $r^2$ according to the deprojected separation of the disk to star in the center of the image.
The scaling is corrected for an inclination of 21\degr\ and a position angle of 76\degr, as derived in Sect.~\ref{subsubsec:outer_ring}.
In each image north is up and east is left.
An arbitrary linear color scale is applied, which is normalized to the maximum flux in each frame. 
Images \textbf{a} and \textbf{b} show the results obtained with SPHERE in CI mode applying a broad $H$- and \textit{K}$_\textit{s}$-band filter, respectively. 
The stellar point spread function was reconstructed by a fit of 20 principal components obtained from a library of reference stars.
In the bottom panel we present the results obtained from SPHERE DPI data in $H$ band. 
Images \textbf{c} and \textbf{d} show the Stokes $Q$ and $U$ parameters for linear polarization, respectively. 
Both polarimetric results are smoothed with a Gaussian kernel having a FWHM of $\sim$50\,mas. This is equivalent to the theoretical SPHERE FWHM in $H$ band.
}
\label{fig:WRAY15788_result}
\end{figure*}

\section{Observational results}
\label{sec:results}

The results of our data reduction are presented in Fig. \ref{fig:WRAY15788_result}.
For both CI and DPI, an artificial mask with a diameter of $\sim$196\,mas is applied to hide the innermost parts of the images that are obscured by the coronagraphic mask and polluted by leaking flux around it.
Furthermore, each pixel is scaled by the squared, deprojected radial separation to the image center to account for intensity loss in scattered light and to highlight features of the disk.
For a correct deprojection we use an inclination of 21\degr\ and a position angle of 76\degr\ following our disk fitting results presented in Sect.~\ref{subsubsec:outer_ring}.

\subsection{CI data}
\label{subsec:results_ci}

\begin{figure}
\resizebox{\hsize}{!}{\includegraphics{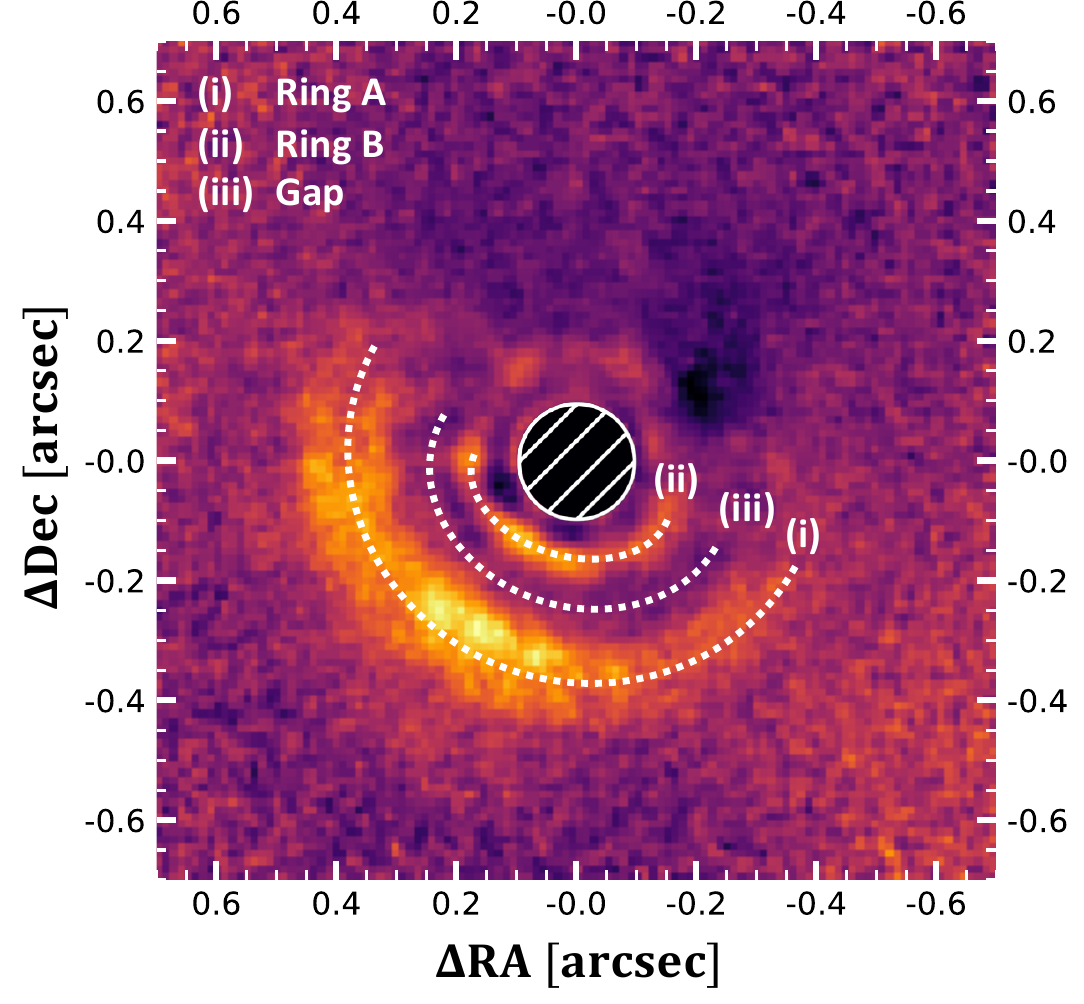}}
\caption{
Annotated version of Fig.~\ref{fig:WRAY15788_result}~\textbf{b}. The discussed features of the disk are highlighted.
}
\label{fig:WRAY15788_result_annotated}
\end{figure}

In frame \textbf{a} and \textbf{b} of Fig. \ref{fig:WRAY15788_result} we present the SPHERE/CI results in $H$ and \textit{K}$_\textit{s}$ band, respectively.
We modeled the stellar PSFs with 20 PCs\footnote{We optimized the number of fitted principal components in order to achieve the best contrast inside the possible disk gap at a projected separation of $\sim$220\,mas.} from our reference library and subtracted these models afterwards.
A bright disk that shows several features is detected in both filters.
The most prominent are the following:
\begin{enumerate}[(i)]
\item \emph{Ring A}: a bright outer arc at an average projected separation of $\sim$370\,mas that is brightest southeast of the star and indistinguishable from background noise in the northwest;
\item \emph{Ring B}: a tentative circular inner ring at an average projected separation of $\sim$170\,mas;
\item a gap in between the two rings.
\end{enumerate}
An annotated image of the disk, in which the main features are highlighted, is presented in Fig. \ref{fig:WRAY15788_result_annotated}.
All these detected features of the disk are analyzed in depth in Sect.~\ref{subsec:imaging_data}, and discussed in Sect.~\ref{sec:discussion}.

\subsection{DPI data}
\label{subsec:results_dpi}

In frames \textbf{c} and \textbf{d} of Fig. \ref{fig:WRAY15788_result} we present Stokes $Q$ and $U$ parameters of the SPHERE/DPI data.
To increase the signal-to-noise ratio in the poor quality observations, we smoothed the images with a Gaussian kernel that has a full width at half maximum (FWHM) of $\sim$50 mas. 
This corresponds to the diffraction limited size of the SPHERE PSF in $H$ band.
Both polarimetric results reveal a strong butterfly-like pattern, approximately centered at the star's position behind the coronagraph.
This agrees with what we expect of azimuthal linear polarization of light scattered by a circumstellar disk.
The positive flux extends down to the artificial mask that we have applied in the image center.
However, this does not necessarily mean that we receive scattered light flux from all separations down to the mask's radial separation of $\sim$98\,mas, due to the poor weather conditions and the previously performed smoothing.
Furthermore, an excess of flux in the southeastern part compared to the northwestern part of the disk is detected in the $Q$ and $U$ images, which agrees very well with the shape of \emph{ring A} that we detect in the CI results.
Moreover, the scattered light flux in the DPI result seems to extend farther out compared to the distinct shape of \emph{ring A} in the CI results.
Whether this extended structure is real or just caused by the applied smoothing and due to the poor weather conditions during the observation will be analyzed in Sect.~\ref{subsubsec:outer_ring}.

\section{Analysis}
\label{sec:analysis}

\subsection{Association of \object{Wray~15-788} with \object{HD~98363}}
\label{subsec:binary}

In our investigation we discovered that \object{Wray~15-788} is part of a multiple system with the A2V star \object{HD~98363} (\object{HIP~55188}).
\object{HD~98363} is a main sequence star of spectral type A2V \citep{Houk75}, and \citet{deZeeuw1999} had identified it as a member of LCC based on Hipparcos astrometry.
\citet{Tetzlaff2011} estimated an isochronal age of 13.0\,$\pm$\,3.7\,Myr to \object{HD~98363} and constrained a mass of 2.0\,$\pm$\,0.1\,$M_\sun$.
Considering binarity with \object{Wray-15-788}, our aim is to derive new estimates for these parameters.
All the important stellar properties are listed in Table~\ref{tbl:WRAY15788_parameters}.

Our companionship analysis is based on parallaxes and proper motions from Gaia DR2 \citep[ICRS, epoch 2015.5,][]{GAIA2018}, which are listed in Table \ref{tbl:WRAY15788_parameters} as well.
The calculated separation of the binary is 49\farcs64974\,$\pm$\,0.05\,mas and the distances agree within 1.08\,$\pm$\,0.84~parsec, statistically consistent with these two stars being co-distant.
The differential velocity in the plane of the sky between the two stars is 0.623\,$\pm$\,0.462\,mas/year.
An estimate of the orbital period of the binary with a separation of 6900\,au is around 330 kyr, with a circular orbital velocity of $0.63$ km/s, which is 0.001 mas/year.
This is marginally consistent with the differential velocity of the two stars above.
So, \object{Wray~15-788} is actually \object{HD~98363~B}: a stellar companion to \object{HD~98363}.

Furthermore, \citet{Chen2012} detected a debris disk around \object{HD~98363} based on 24\,$\mu$m and 70\,$\mu$m photometry from \emph{Spitzer} MIPS \citep{Werner2004,Rieke2004}.
\citet{Moor2017} reported a non-detection of CO with an upper limit of 0.036\,Jy\,km/s on the integrated line flux of \element[][12]{CO} J=2--1.
This gas-poor debris disk around \object{HD~98363} is especially interesting due to our finding of a disk around \object{Wray~15-788}.
A discussion of this special binary system with two hosts of circumstellar disks is presented in Sect.~\ref{subsec:comparison_hd98636}.

\begin{figure}
\resizebox{\hsize}{!}{\includegraphics{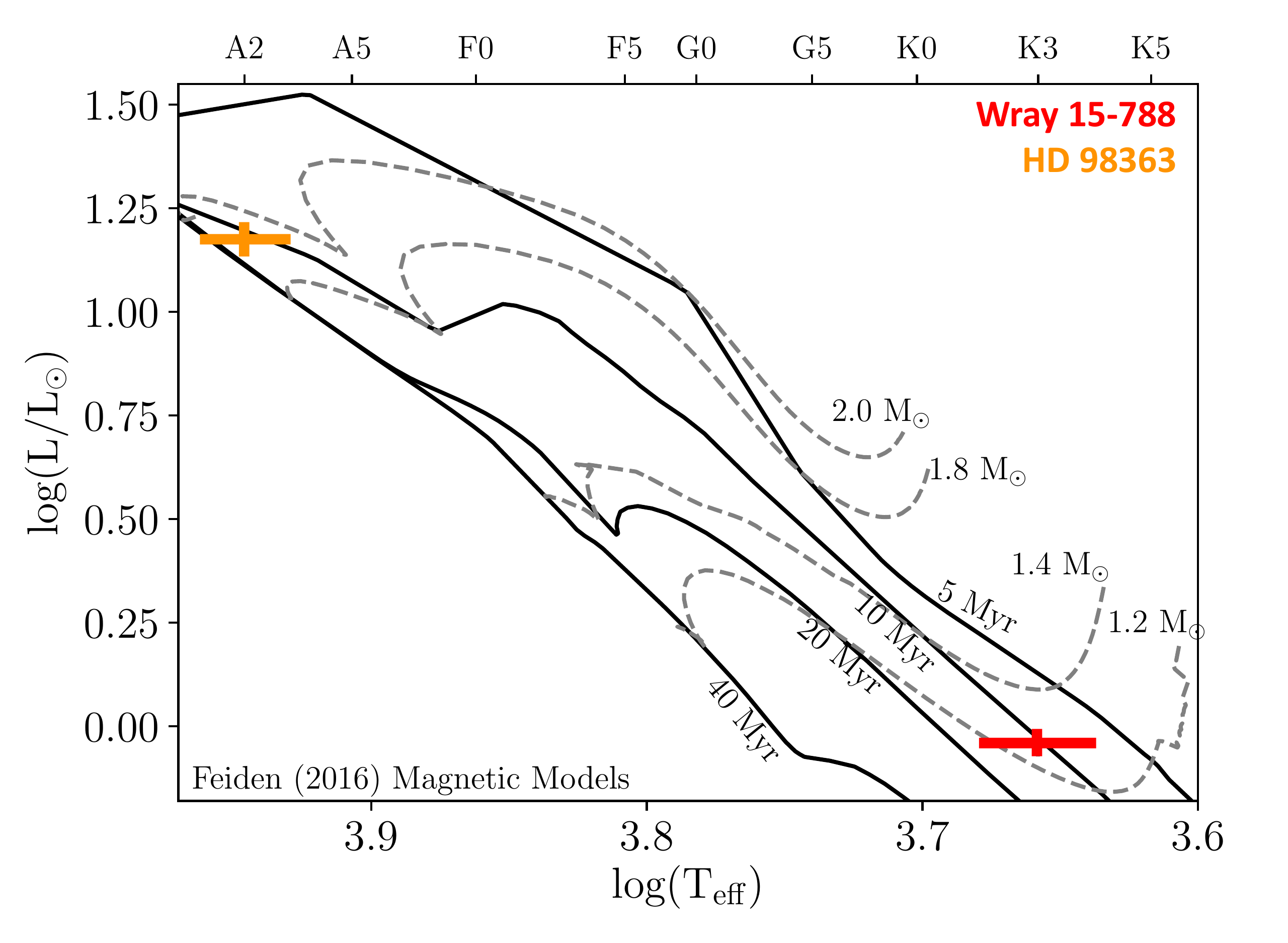}}
\caption{
Hertzsprung-Russell diagram for the binary system of \object{Wray~15-788} and \object{HD~98363}. 
We plot pre-main sequence tracks and isochrones according to \citet{Feiden16} to constrain masses and ages of the companions.
}
\label{fig:hrd}
\end{figure}

To derive consistent masses and ages of the binary system we analyzed the two stars within an Hertzsprung-Russell diagram, as presented in Fig.~\ref{fig:hrd}.
We estimated the masses and ages using Feiden/Dartmouth tracks \citep{Feiden16}.
These models include magnetism below 1.7\,$M_\sun$ and yield a consistent age for the Upper Scorpius subgroup of Sco-Cen \citep{Feiden16}.
Therefore, they define a good basis for an analysis of our two LCC objects.
Using a flat age and \citet{Maschberger2013} initial mass function as priors, we obtain masses of $1.26^{+0.07}_{-0.22}\,M_\sun$ and $1.92^{+0.08}_{-0.08}\,M_\sun$ for \object{Wray~15-788} and \object{HD~98363}, respectively.
Furthermore, we obtain an age of $11^{+16}_{-7}$\,Myr for the two companions, apparent in Fig.~\ref{fig:hrd}.

\subsection{SED modeling}
\label{subsec:sed_modeling}

\begin{figure}
\resizebox{\hsize}{!}{\includegraphics{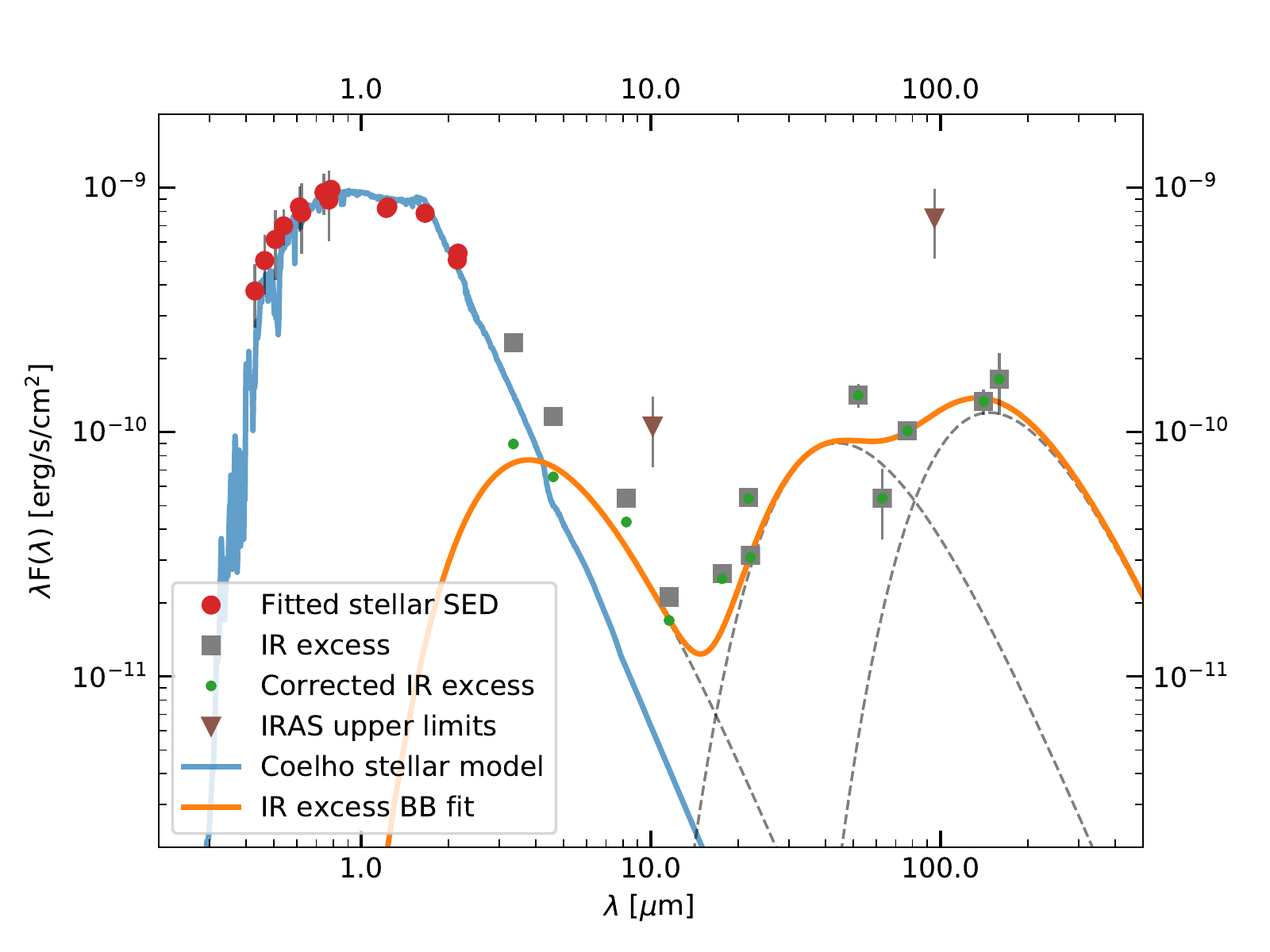}}
\caption{
De-reddened spectral energy distribution of \object{Wray~15-788}.
The blue curve shows a Coelho stellar model \citep{Coelho2014} with $T_\mathrm{eff}$\,=\,4250\,K, $\log(g)$\,=\,4.5, [Fe/H]\,=0, [$\alpha$/Fe]=0, and $A_V$\,=\,0.74 that is fitted to the red data points from the \textit{APASS}, \textit{Gaia}, \textit{2MASS}, and \textit{DENIS} photometry.
The gray squares denote flux measurements from \textit{WISE}, \textit{IRAS}, \textit{AKARI/FIS}, and \textit{AKARI/IRC,} and the brown triangles provide upper limits from \textit{IRAS}.
Three blackbodies with $T_\mathrm{dust, 1}$\,=\,969\,K, $T_\mathrm{dust, 2}$\,=\,83\,K, and $T_\mathrm{dust, 3}$\,=\,25\,K were simultaneously fitted to the green data points that denote the object's far-infrared excess, corrected for stellar contamination.
The individual blackbody functions are indicated by the dashed gray lines, whereas their sum is presented by the solid orange curve.
}

\label{fig:wray_15788_sed}
\end{figure}

We obtained the available SED of \object{Wray~15-788} presented in Fig.~\ref{fig:wray_15788_sed}.
It is rather well sampled by photometry from \textit{APASS} \citep{Henden2014}, \textit{Gaia} \citep{GAIA2018}, \textit{2MASS} \citep{Cutri2012a}, and \textit{DENIS} \citep{Epchtein1997} for wavelengths up to approximately 2\,$\mu$m; however, we only have a few data points from \textit{WISE} \citep{Cutri2012b}, \textit{IRAS} \citep{Neugebauer1984}, \textit{AKARI/FIS} \citep{Murakami2007,Kawada2007}, and \textit{AKARI/IRC} \citep{Murakami2007,Ishihara2010}, and additional upper limits from \textit{IRAS} for wavelengths longer than this.
In particular, there is no data available beyond 160\,$\mu$m.

To evaluate whether the system is a potential gas-rich protoplanetary disk, we aimed to derive the fractional infrared luminosity
\begin{align}
f =\frac{L_\mathrm{IR}}{L_{*}}\;,
\end{align}
where $L_\mathrm{IR}$ and $L_{*}$ denote the bolometric luminosities of the infrared excess and the star, respectively.
To get accurate estimates of both bolometric luminosities, we fitted the stellar spectrum and the infrared contribution to the SED due to circumstellar material individually with a suitable model.

Analyzing the SED with VOSA \citep{Bayo2008} indicates an infrared excess for wavelengths longer than $W1$ ($\lambda_\text{c}^{W1}$=3.35\,$\mu$m).
Thus, we only used the data points at wavelengths shorter than this to fit the spectrum of the star.
For this purpose we applied a Coelho stellar model \citep{Coelho2014} that depends on effective temperature $T_\mathrm{eff}$, surface gravity $\log(g)$, and the metallicity parameters [Fe/H] and [$\alpha$/Fe] of the star.
Furthermore, we assumed a total extinction $A_V$ in the range of 0.5\,mag\,<\,$A_V$\,<1.5\,mag, in agreement with $A_V$\,=\,0.88\,$\pm$\,0.18\,mag as determined by \citet{Pecaut2016} for \object{Wray~15-788}.
The $\chi^2$ fit yields a template stellar spectrum with $T_\mathrm{eff}$\,=\,4250\,K, $\log(g)$\,=\,4.5, [Fe/H]\,=\,0, [$\alpha$/Fe]\,=\,0, and $A_V$\,=\,0.74, which is represented by the blue line in Fig.~\ref{fig:wray_15788_sed}.
These model parameters agree very well with the stellar properties of \object{Wray~15-788} determined by \citet{Pecaut2016} and within the scope of this work as presented in Sect.~\ref{subsec:binary}.
The red points in Fig.~\ref{fig:wray_15788_sed} show the fitted photometric data points from \textit{APASS}, \textit{Gaia}, \textit{2MASS}, and \textit{DENIS}, for which a de-reddening according to the best fit stellar model was applied.
The gray squares and brown triangles represent the infrared flux of the system and upper limits to it, respectively.

To determine $L_\mathrm{IR}$ we focused on the excess at wavelengths longer than 2\,$\mu$m.
First, we corrected the available data points for the contamination by stellar flux using the best fit Coelho stellar model that we had found for the star.
The corrected data is presented by the green dots in Fig.~\ref{fig:wray_15788_sed}.
To model the infrared SED of circumstellar material around \object{Wray~15-788} we used three individual blackbodies: one to account for a hot, inner component at wavelengths between 2\,$\mu$m and 10\,$\mu$m, and two additional blackbodies to characterize the colder, outer parts of the disk.
The corrected data was fitted by the sum of these blackbody functions using a Levenberg–Marquardt non-linear least-squares solver \citep{Levenberg1944,Marquardt1963} and taking the inverse of each data point's uncertainty as corresponding numerical weight.
This yields a best fit result with effective blackbody temperatures of 969\,K, 83\,K, and 25\,K.
The individual blackbodies are indicated by the gray dashed lines in Fig.~\ref{fig:wray_15788_sed} and their sum is represented by the solid orange curve.

Integrating the stellar model and the fit of the infrared excess over the entire spectral range yields a fractional infrared luminosity of $f $\,\ga\,$0.27$.
Due to the incomplete SED for wavelengths longer than 160\,$\mu$m this value has to be interpreted as a lower threshold.

\subsection{Imaging data}
\label{subsec:imaging_data}

Both classical and dual-polarimetric imaging results confirm a resolved, asymmetrical, disk-like structure around \object{Wray~15-788}.
Consequently, we tried to quantify the reliability of the features detected in Fig. \ref{fig:WRAY15788_result}.

\subsubsection{Ring A}
\label{subsubsec:outer_ring}

\emph{Ring A} of the disk is detected with SPHERE/CI in the $H$ and \textit{K}$_\textit{s}$ bands, and the butterfly patterns in Stokes $Q$ and $U$ frames from SPHERE/DPI are a strong confirmation of a scattering, disk-like structure around \object{Wray~15-788}.
Due to the higher signal-to-noise ratio of the disk detection, we restricted our subsequent analyses to the the \textit{K}$_\textit{s}$-band data.

\subsubsection*{Disk fitting}

To determine the inclination of the disk, we fitted \emph{ring A} by an elliptical aperture.
For this purpose, we used the SPHERE/CI \textit{K}$_\textit{s}$-band result (see Fig. \ref{fig:WRAY15788_result}~\textbf{b}).

We smoothed the images with a Gaussian kernel having a FWHM of 55\,mas, which corresponds to the theoretical size of the instrument's PSF in \textit{K}$_\textit{s}$ band.
To focus the fit only on the actual signal of the disk, an inner and outer mask were placed around \emph{ring A}.
The mask's inner and outer radii were set to 0\farcs31 and 0\farcs47, respectively.
Afterwards, we split the image in 100 azimuthal slices, centered at the star's position.
Within each slice, we determined the pixel of maximum flux.
In order to reject background signal we set a lower threshold that corresponds to the median flux at the pixel's separation to the star.
Finally, an ellipse was fitted to the remaining pixels of maximum flux by a linear least-squares algorithm according to the implementation of \citet{Fitzgibbon1999}.
We used a model of an arbitrary, two-dimensional ellipse with five free parameters $\delta x$, $\delta y$, $a$, $b$, $\varphi$.
The meaning of these parameters is explained in Table \ref{tbl:ellipse_parameters}.
\begin{table}
\caption{Ellipse parameters.}
\label{tbl:ellipse_parameters}
\def\arraystretch{1.2}
\setlength{\tabcolsep}{12pt}
\centering
\begin{tabular}{@{}ll@{}}
\hline\hline
Parameter & Explanation \\ 
\hline
$(\delta x, \delta y)$ & Center offset from the star position \\
$a$ & Semimajor axis \\
$b$ & Semiminor axis \\
$\varphi$ & Position angle of the semimajor axis\\
\hline
\end{tabular}
\end{table}
The disk inclination $i$ can consequently be calculated as
\begin{align}
i = \arccos\left(\frac{b}{a}\right)\,.
\end{align}

To assess an estimate of the uncertainties on our best fit parameters we assumed that the locations of the initial positions used for the fit are uncertain to the FWHM that was applied for smoothing.
Therefore, we randomly sampled the initial positions around the previously used values within a box with the size of the FWHM.
We used a flat prior in the $x$ and $y$ directions, and repeated the fitting procedure $10^6$ times.
We obtain symmetric posterior distributions of the ellipse parameters and use the standard deviation as an estimate for the statistical uncertainties of the fit parameters.

The best fit values and corresponding uncertainties of the ellipse parameters are presented in Table \ref{tbl:ellipse_best_fit}.
\begin{table}
\caption{Best fit parameters of the ellipse.}
\label{tbl:ellipse_best_fit}
\def\arraystretch{1.2}
\setlength{\tabcolsep}{12pt}
\centering
\begin{tabular}{@{}lc@{}}
\hline\hline
Parameter & Best fit value \\ 
\hline
$\delta x$ [pix]\tablefootmark{a} & $0.96\pm1.17$ \\
$\delta y$ [pix] & $1.12\pm2.01$ \\
$a$ [pix] & $32.56\pm0.81$ \\
$b$ [pix] & $30.39\pm2.12$ \\
$\varphi$ [\degr] & $76\pm16$ \\
$i$ [\degr] & $21\pm6$ \\
\hline
\end{tabular}
\tablefoot{
\tablefoottext{a}{To convert pixels to projected separations in mas the results must be multiplied with the pixel scale of the detector, which is 12.265\,$\pm$\,0.009\,mas~per~pixel in \textit{K}$_\textit{s}$ band.}
}
\end{table}
The fitting yields a disk inclination of $i$\,=\,21\degr$\,\pm\,$6\degr\ and a position angle of $\varphi$\,=\,76\degr$\,\pm\,$16\degr.
These constraints are rather loose due to the low inclination of the system and because the data points used for the fit sample less than half of an ellipse.
Additional high-quality data is required to confine this parameter space.

\subsubsection*{Disk signal-to-noise ratio estimation}

\begin{figure}
\resizebox{\hsize}{!}{\includegraphics{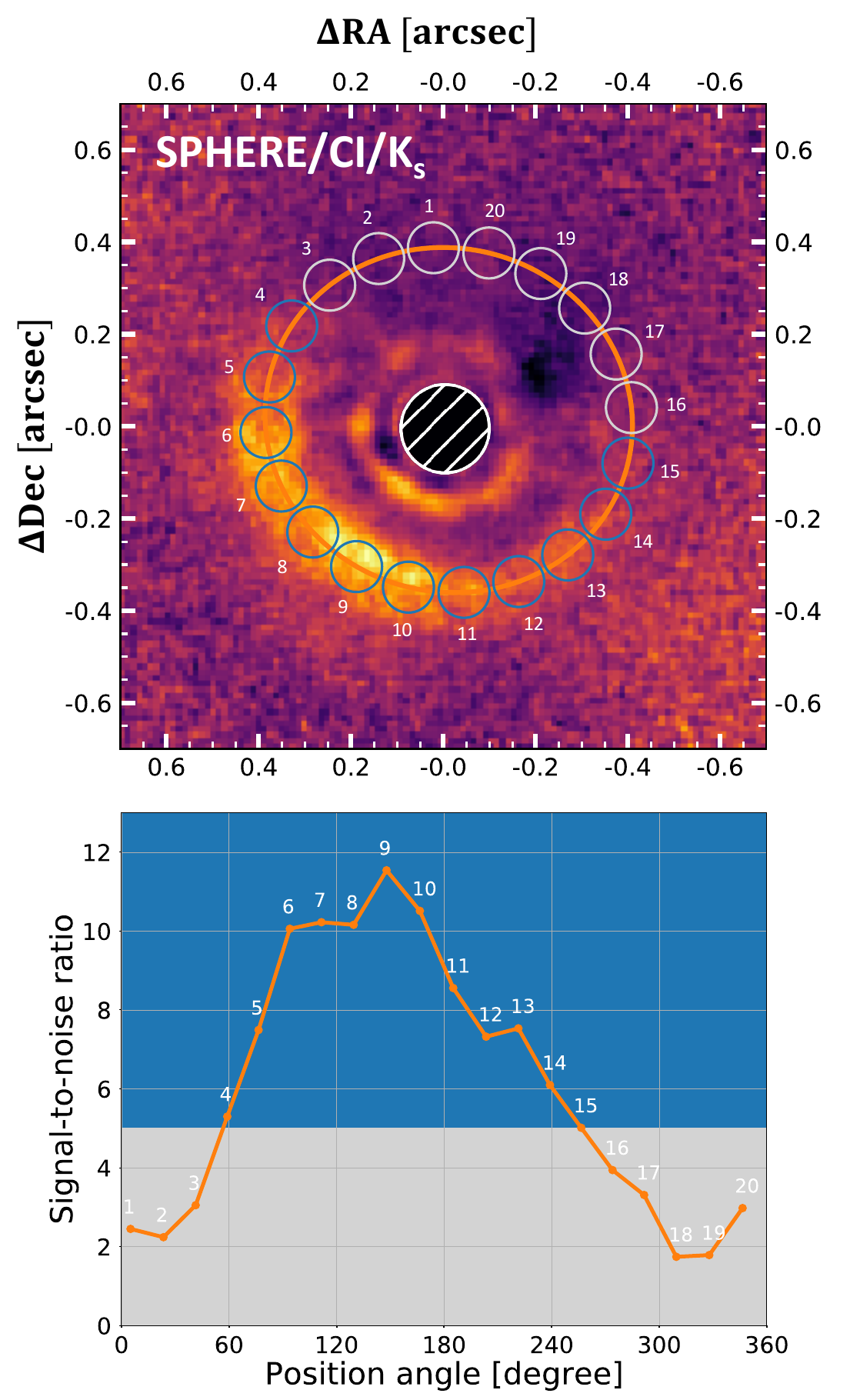}}
\caption{
\emph{Top}: 
Signal-to-noise ratio measurements in circular apertures around the best fit ellipse (orange line) for the \textit{K}$_\textit{s}$-band data from Sect.~\ref{subsubsec:outer_ring}. 
The blue apertures contain flux of \emph{ring A} according to the applied 5$\sigma$ criterion, whereas the gray apertures reside within the background dominated regime.
\emph{Bottom}:
Signal-to-noise ratio within the circular apertures from the top panel, sorted by position angle.
The gray and blue areas mark the regimes below and above the 5$\sigma$ threshold to distinguish background and disk apertures, respectively.
}
\label{fig:snr_measurements}
\end{figure}

\begin{figure*}
\centering
\includegraphics[width=17cm]{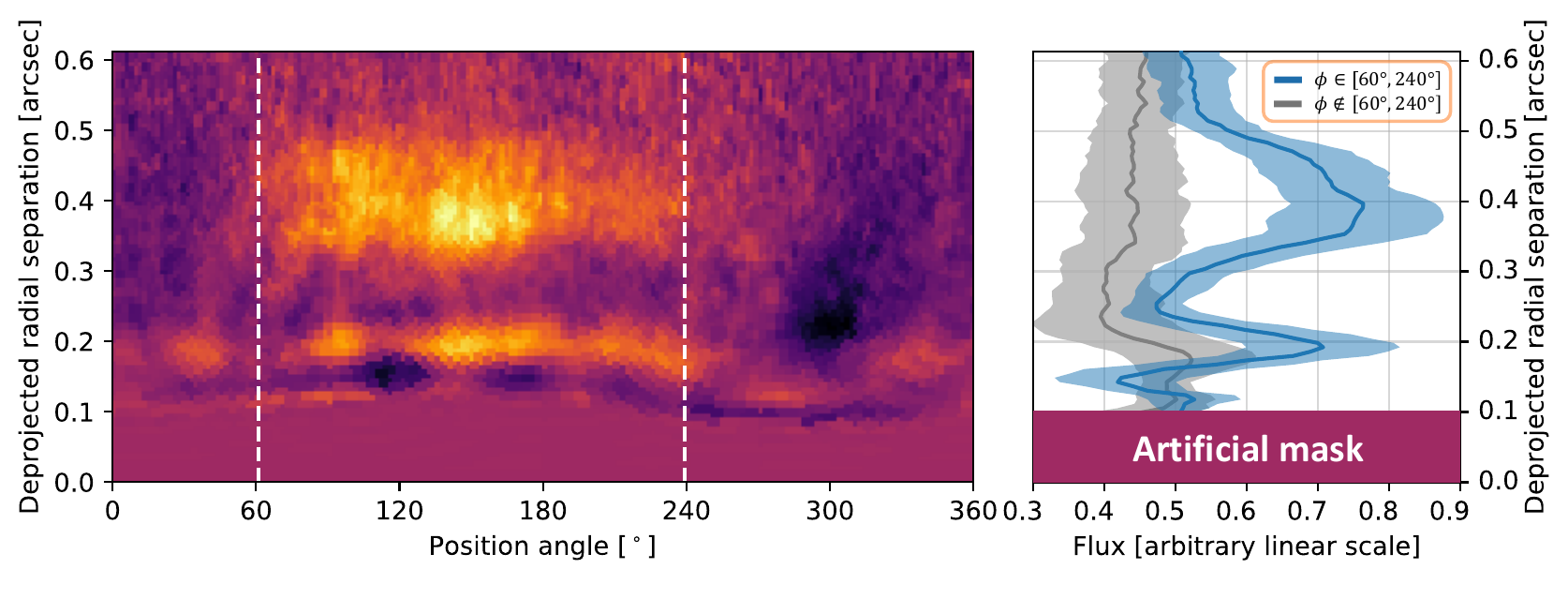}
\caption{\emph{Left panel}: Polar projection of the SPHERE/CI result in \textit{K}$_\textit{s}$ band. The image is corrected for the offset and the inclination of the best fit ellipse to \emph{ring A}. The white dashed lines indicate the range of position angles in which we detect scattered light flux from \emph{ring A} with a signal-to-noise ratio higher than 5. \emph{Right panel}: Radial brightness profile of the disk and background noise. The blue curve shows the averaged disk signal for position angles in the range of 60\degr\,$\leq\,\phi\,\leq$\,240\degr\ and the gray curve represents the average noise over all remaining position angles. The envelopes indicate the corresponding standard deviations.}
\label{fig:WRAY_15788_polar_map}
\end{figure*}

As presented in Fig.~\ref{fig:WRAY15788_result}, the azimuthal brightness profile of \emph{ring A} varies widely.
Starting north of the star, the disk flux increases with increasing position angle $\phi$.
The maximum intensity of this ring is located southeast of the star.
Thereafter, the flux decreases with increasing position angle until the disk signal cannot be distinguished from the background noise.

We aimed to determine a range of position angles in which we have a significant detection of scattered light flux from \emph{ring A}.
Therefore, we used the best fit ellipse that we had derived earlier and distributed evenly spaced circular apertures along it, as indicated in Fig. \ref{fig:snr_measurements}.
We measured the mean flux and standard deviation inside each individual circular aperture.
The average flux values $\mu_i$ provide an estimate of the signal at the position angle of the corresponding aperture.
To get an estimate of the background noise, we performed a sigma clipping on the array of aperture fluxes.
For the clipping we ran five iterations with no lower threshold and an upper threshold of 1$\sigma$ to exclude strong contamination by disk flux.
After this selection, we calculated the average of the remaining standard deviations for estimating the background noise $\sigma_\text{bg}$.
The signal-to noise ratio (S/N) of each individual aperture is calculated as
\begin{align}
\text{(S/N)}_i = \frac{\mu_i}{\sigma_\text{bg}}.
\end{align}
We applied an arbitrary threshold of (S/N)$_i$\,>\,5 for the selection of disk apertures and rejection of background signal.
This selection criterion, however, agrees very well with the range of position angles, where the disk signal can still be distinguished from background noise by visual inspection (see Fig.~\ref{fig:WRAY15788_result}~b).

The bottom panel of Fig.~\ref{fig:snr_measurements} shows the measured S/N inside each aperture and compares the values to the applied threshold criterion.
Data points in the blue regime of the plot refer to apertures above the threshold, and are therefore considered to indicate a detection of scattered light flux from \emph{ring A}.
The gray regime, however, represents apertures that are dominated by background noise.
This color scheme coincides with the colors chosen for the circular apertures in the top panel of the figure. 
Considering these blue apertures, we derive a range of 60\degr\,\la\,$\phi$\,\la\,240\degr\ in which we are confident at the 5$\sigma$ level to detect scattered light flux of the disk.

Furthermore, we create an inclination-corrected polar projection of the SPHERE/CI results in \textit{K}$_\textit{s}$ band as presented in the left panel of Fig. \ref{fig:WRAY_15788_polar_map}.
Averaging over the position angles within the derived range of 60\degr\,\la\,$\phi$\,\la\,240\degr\ (white dashed lines) yields the radial brightness profile presented by the blue curve in the right panel of Fig. \ref{fig:WRAY_15788_polar_map}.
The gray curve presents the average over the remaining range of position angles in which we do not detect significant disk signal.
From these profiles, it becomes clear that we resolve both the gap and \emph{ring A}.
The latter even shows some hints for substructures as the averaged flux does not decrease as steeply in the radially outward direction as it does towards the inward gap.
Even beyond deprojected separations of 0\farcs5 the average flux of the disk signal is significantly higher than the average background noise.
This is a strong confirmation for scattering material beyond the sharp edge of \emph{ring A}, which was already implied by the DPI data presented in Fig.~\ref{fig:WRAY15788_result}~\textbf{c} and \textbf{d}.

Additionally, the polar deprojection allowed us to estimate physical separations of the disk features that we have detected:
\emph{Ring A} has its peak of scattered light intensity at $\sim$56\,au, the scattered light flux is lowest inside the gap at $\sim$35\,au, and \emph{ring B} has a separation of $\sim$28\,au.

\subsubsection{Ring B}
\label{subsubsec:inner_ring}

To quantify the significance of the detected inner substructure from the SPHERE imagery, we investigated the polar projection presented in the left panel of Fig. \ref{fig:WRAY_15788_polar_map}.
Between the two white dashed lines at a deprojected separation of $\sim$0\farcs2, \emph{ring B} appears to be partly parallel to the resolved \emph{ring A} and even has a similar azimuthal brightness distribution.
Therefore, it is possible that we detect parts of an inner substructure with similar scattering properties.
In the remaining range of position angles, however, the flux received from \emph{ring B} is significantly smaller and its deprojected radial separation varies strongly.
This is an indication for a symmetrical, probably non-astrophysical residual around the coronagraph that gets distorted by the inclination correction that we perform to create Fig. \ref{fig:WRAY_15788_polar_map}.

\begin{figure}
\resizebox{\hsize}{!}{\includegraphics{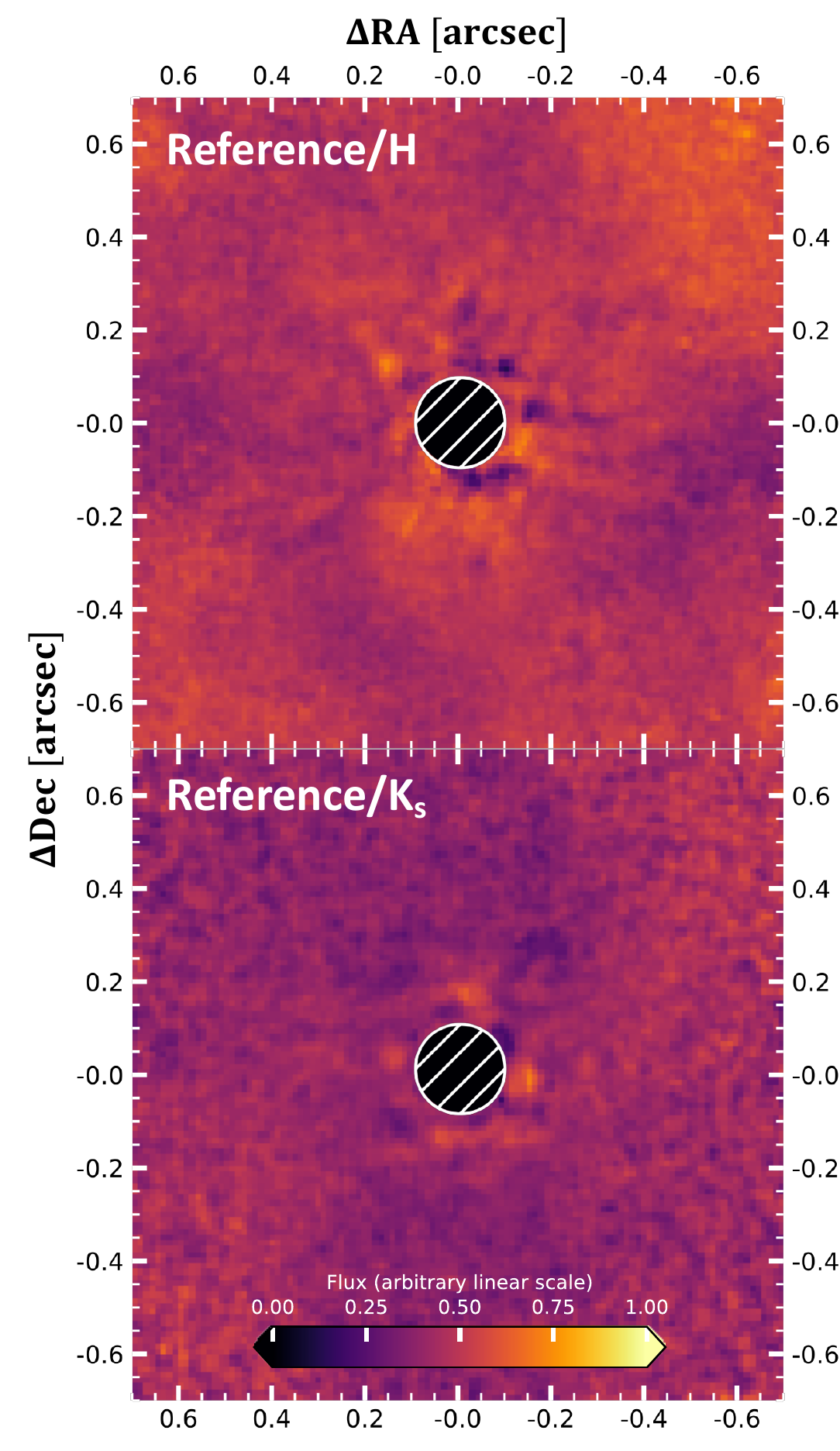}}
\caption{
Comparison to average images obtained from reduced reference library targets in \textit{H} band (\emph{top}) and \textit{K}$_\textit{s}$ band (\emph{bottom}).
}
\label{fig:reference_library_median_image}
\end{figure}

To test this hypothesis, we compared our result to data from our reference library, obtained with the same observational setup.
These data were reduced analogously to the approach we describe in Sect.~\ref{sec:data_reduction}.
For each target we applied RDI in combination with PCA and we fitted 20 components for modeling the stellar PSF.
All residuals were averaged individually for both filters and to enhance the comparability to our previous results from Fig. \ref{fig:WRAY15788_result}, we applied the same radial scaling and masking of the innermost region.
However, we did not perform any de-rotation of the images.
Because all data was obtained in pupil stabilized mode, this approach ensures proper alignment of potential instrumental artifacts.
These reference images in \textit{H} and \textit{K}$_\textit{s}$ band are presented in Fig.~\ref{fig:reference_library_median_image}.

\begin{figure}
\resizebox{\hsize}{!}{\includegraphics{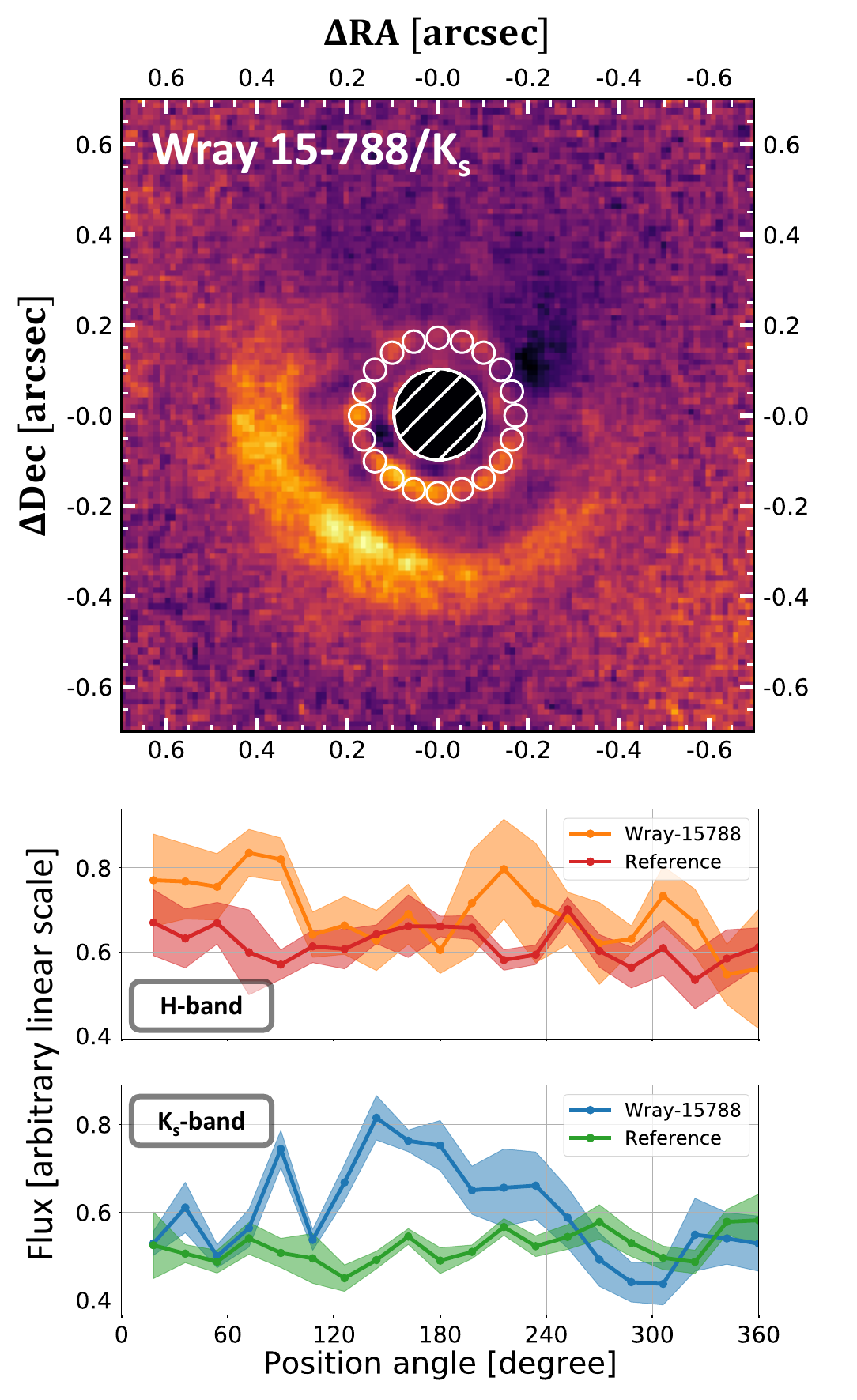}}
\caption{
Comparison to average images obtained from reduced reference library targets.
\emph{Top}: 
Locations of the flux apertures that are distributed alongside \emph{ring B} (white circles).
\emph{Bottom}:
Flux measurements in the apertures from the top panel as a function of position angle.
The solid lines correspond to the average flux per aperture.
The envelope indicates the corresponding standard deviation.
}
\label{fig:reference_library_flux_inner_ring}
\end{figure}

We detect some features close to the coronagraph in both reference images.
The \textit{H}-band data shows a rather unstructured speckle pattern similar to the science result in that filter (compare to Fig.~\ref{fig:WRAY15788_result}~\textbf{a}), while the \textit{K}$_\textit{s}$-band reference residuals reveal a faint inner ring at the same projected radial separation of $\sim$170\,mas, but \emph{ring B} that we detect around \object{Wray~15-788} is significantly brighter in the southeast than the residuals from the reference library.
Northwest of the star, however, the intensity of \emph{ring B} is equal for \object{Wray~15-788} and the reference stars. 
This is consolidating our claim that we actually detect the scattered light flux of an inner substructure southwest of the star.
To quantify this observation, we performed photometry in circular apertures distributed alongside \emph{ring B} as indicated in the top panel of Fig.~\ref{fig:reference_library_flux_inner_ring}.
We chose a radial separation of $\sim$170\,mas to the star and each aperture has a radius of $\sim$25\,mas, which corresponds to the measured width of \emph{ring B}.
These measurements were performed for both filter combinations and the corresponding reference results.
We determine the average flux and standard deviation per aperture and plot these as a function of position angle as presented in the bottom panel of Fig.~\ref{fig:reference_library_flux_inner_ring}.
In \textit{H} band no strong differences between the flux around \object{Wray~15-788} and the reference image can be detected; instead, we observe a significant peak in \textit{K}$_\textit{s}$-band.
Within a range of position angles of 120\degr\,\la\,$\phi$\,\la\,240\degr\ the flux measured in the apertures on \emph{ring B} around \object{Wray~15-788} is greater than the flux from the reference image within the same range of position angles.
The determined angular interval lies within the interval where we detect \emph{ring A} with a S/N greater than 5.
This strengthens the claim that we actually detect parts of a inner substructure around \object{Wray~15-788}.
Because we do not spatially resolve these structures, we cannot make an accurate estimate of its inclination.

Even though Fig.~\ref{fig:WRAY_15788_polar_map} implies the detection of another gap interior to \emph{ring B}, we do not trust this feature, because it is placed very close to the inner working angle (IWA) of the coronagraph of 100\,mas (Wilby et al. in prep.).
For this reason we consider it to be an artifact caused by our post-processing strategy.

\subsubsection{Gap}
\label{subsubsec:gap}

For the CI results we detect a significant decrease in flux interior to \emph{ring A}.
Depending on the position angle, this radial gradient is steepest at a projected separation of $\sim$250\,mas.
We do not recover this drop in scattered light surface brightness from the polarimetric dataset, but there are several factors that can explain this behavior (e.g., non-optimal weather conditions or smoothing with a Gaussian kernel).
Furthermore, we can conclude from the polar deprojection of the disk in Fig. \ref{fig:WRAY_15788_polar_map} that we are able to spatially resolve this radial drop in intensity.
Because we detect this decrease in scattered light flux even in data processed without proper subtraction of the stellar PSF by RDI+PCA (see Appendix \ref{subsec:other_reductions_ci}), we conclude that it is a real phenomenon. 
Possible explanations for this very certain dip in surface brightness are either a shadowed region or a physical cavity within the disk.

\subsubsection{Detection limits}
\label{subsubsec:detection_limits}

\begin{figure}
\resizebox{\hsize}{!}{\includegraphics{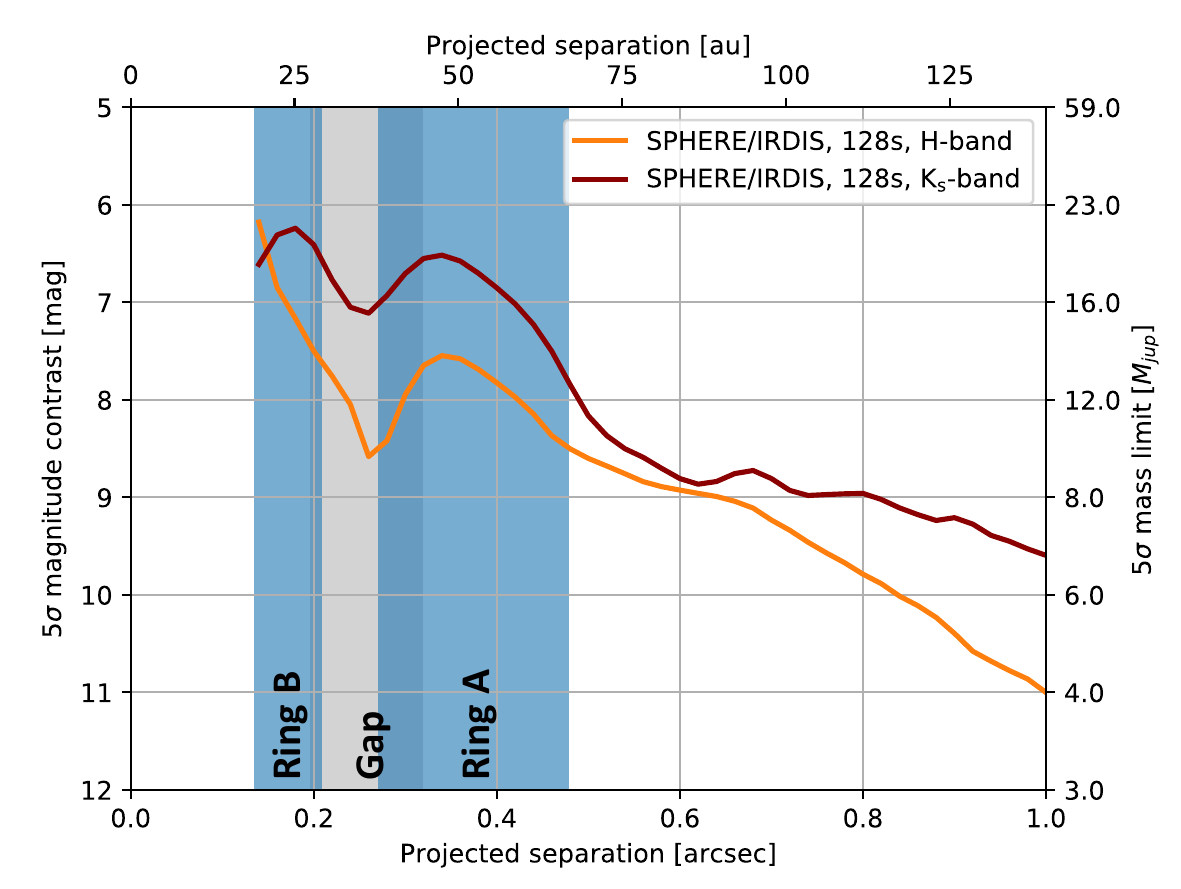}}
\caption{
Magnitude and mass limits (5$\sigma$) up to 1\arcsec around \object{Wray~15-788}.
The positions of the two rings and the gap are indicated in blue and gray, respectively.
The magnitude contrast was converted to an upper mass limit by AMES-Cond models for 11\,Myr old objects.
Inside the gap we are sensitive to companions as massive as 10\,$M_\text{jup}$ (\textit{H} band) and 15\,$M_\text{jup}$ (\textit{K}$_\textit{s}$ band).
}
\label{fig:detection_limits}
\end{figure}

To derive mass limits of an undetected companion to \object{Wray~15-788}, we calculated 5$\sigma$ contrast curves using the standard routine of the PynPoint package \citep{Stolker2019}. 
Artificial companions were obtained from the non-coronagraphic flux images that we had taken alongside our science observations.
They were scaled to correct for the difference in exposure times and the attenuation of a neutral density filter.
The injection was performed for six evenly spaced azimuthal directions and radial separations ranging from 0\farcs15 to 1\arcsec with a step size of 20\,mas.
We present the 5$\sigma$ detection limits for both CI filters in Fig. \ref{fig:detection_limits}.
Close to the star, we applied a correction to account for small sample statistics according to \citet{Mawet2014}.
The conversion from magnitude contrast to a detectable mass threshold was performed using AMES-Cond\footnote{The latest version of these models were obtained from \url{https://www.phoenix.ens-lyon.fr/Grids/AMES-Cond}} atmospheric models of 11\,Myr old substellar objects \citep{Allard2001,Baraffe2003}.
We indicate the position of the two rings and the gap in blue and gray, respectively.
Because the structures are not detected face-on, there are small spatial overlaps between rings and the gap.
At the center of the gap we are sensitive to companions as massive as 10\,$M_\text{jup}$ and 15\,$M_\text{jup}$ in \textit{H} and \textit{K}$_\textit{s}$ band, respectively.
At separations larger than 1\arcsec, we can rule out companions more massive than 4\,$M_\text{jup}$.
We did not apply any correction for reddening and extinction by interstellar matter or disk material in our analysis.

\section{Discussion}
\label{sec:discussion}

\subsection{SED analysis}
\label{subsec:discussion_sed}

\citet{Pecaut2016} classify \object{Wray~15-788} as a potential host of a protoplanetary disk based on two criteria: (i) the H$\alpha$ emission as an indicator of accretion from a gas-rich disk and (ii) the presence of an infrared excess in its SED indicative of dust grains.
The EW(H$\alpha$) threshold for accretion from \citet{Barrado2003} for a K3 star is 4.1\,\AA;
the measured EW(H$\alpha$) from \citet{Pecaut2016} is 10.3\,\AA.
The full width at 10\% max of the line is 430 km/s (Pecaut, private communication 2018), which exceeds the empirical criterion for accretion of 270 km/s \citep{White2003}, and thus is consistent with ongoing accretion.
Furthermore, \citet{Pecaut2016} derive the extent of the infrared excess by determining the $K_s$--$W3$ and $K_s$--$W4$ colors from \emph{2MASS} \citep{Henden2012} and \emph{WISE} \citep{Cutri2012b} magnitudes (see Table~\ref{tbl:WRAY15788_parameters}).
According to the empirical threshold determined by \citet{Luhman2012}, a protoplanetary disk is expected to have excesses exceeding $K_s$--$W3$\,>\,1.5 and $K_s$--$W4$\,>\,3.2.
With $K_s$--$W3$\,=\,1.76\,$\pm$\,0.04 and $K_s$--$W4$\,=\,4.3\,$\pm$\,0.04, \object{Wray~15-788} clearly meets these criteria.

The conclusion that \object{Wray~15-788} hosts a protoplanetary disk is clearly supported by the analysis of the object's SED presented in Sect~\ref{subsec:sed_modeling}.
Comparison of the derived fractional infrared luminosity $f$\,\ga\,0.27 with empirical thresholds of \citet{Dominik2003} and \citet{Lagrange2000}, strongly imposes that \object{Wray~15-788} harbors a gas-rich protoplanetary disk rather than a debris disk where most of the gas has already dissipated.
Usual fractional infrared luminosities of the latter category are in all known cases indeed smaller than $10^{-2}$.
So, \object{Wray~15-788} exceeds this threshold by more than one order of magnitude.

Furthermore, the fit of the flux at wavelengths longer than 2\,$\mu$m as presented in Fig.~\ref{fig:wray_15788_sed} clearly illustrates that the infrared SED of the system cannot be described by a single belt model alone.
The excess at near-infrared wavelengths (2\,$\mu$m\,<\,$\lambda$\,<\,10$\mu$m) modeled by a blackbody with an effective temperature of 969\,K strongly indicates the presence of a hot, inner component of the disk around the star \citep[e.g.,][]{Tilling2012}.
Based on the high effective temperature, this inner component must be located close to the dust sublimation radius, and therefore definitely interior to the IWA of the applied coronagraph.
Thus, we can rule out with high confidence that \emph{ring B}, as potentiality detected in the imaging data, is equivalent to this hot inner component of the disk.
Comparing the SPHERE imagery with the object's SED suggests that \emph{ring A} and \emph{B} are represented by the infrared excess at wavelengths longer than 10\,$\mu$m.

Around 10\,$\mu$m there is an apparent dip in the SED that is followed by a positive gradient towards longer wavelengths.
These characteristics of the SED impose a physical, dust depleted cavity inside the disk that is enclosed by an extended, colder component of disk material.
For these reasons we conclude that \object{Wray~15-788} hosts a protoplanetary disk at a transition stage \citep{Strom1989, Furlan2009}.

\subsection{Disk morphology}
\label{subsec:morphology}

The disk around \object{Wray15-788} appears highly asymmetric with flux only detected on the southeastern side.
Because we can detect this asymmetry in the DPI data as well, and even in the CI data processed without proper subtraction of the stellar PSF, we can rule out that this appearance is an artifact of our post-processing.
These alternative reductions are presented in Appendix \ref{sec:other_reductions}.
Due to the low inclination of the disk, our observation probes only a limited range of scattering angles, which should not be significantly smaller than $\sim$50\degr. 
In this range of scattering angles the scattering phase functions of typical disks are flat \citep[see][for an extensive overview]{Hughes2018}. 
Thus, we should receive scattered light from all azimuthal positions of the disk.
This is indeed true even for slightly more inclined debris and gas-rich disks, such as {\object{HD~181327} \citep[$i$\,$\approx$\,32\degr; e.g.,][]{Soummer2012}, \object{PDS~66} \citep[$i$\,$\approx$\,32\degr; e.g.,][]{Schneider2014,Wolff2016}, \object{V4046~Sgr} \citep[$i$\,$\approx$\,34\degr; e.g.,][]{Rapson2015}, or \object{HD~100453} \citep[$i$\,$\approx$\,38\degr; e.g.,][]{Benisty2017}.
For \object{PDS~66}, \citet{Wolff2016} measure contrasts in scattered light brightness of 2.1 and 1.6 for \textit{H} and \textit{K}1 band, respectively, between the near and far sides of the disk.
Adopting this contrast ratio for \object{Wray~15-788} shows that we should detect the far side of the disk (northwestern part) at a S/N higher than 5, because the near side (southeastern part) is detected at a S/N of approximately 12.
As presented in Fig.~\ref{fig:snr_measurements}, this is obviously not the case.
Although we cannot fully rule out the possibility that the asymmetry is caused by a larger contrast ratio between the near and far sides of the disk so that the S/N in the northwest drops below our detection ability, it seems unlikely that the apparent morphology is caused by scattering phase function effects.

We can thus conclude that the asymmetry is either caused by a strong azimuthal variation in surface density or scale height of the disk, or that a shadow is cast on \emph{ring A} by unresolved disk structures interior to the structures we detect in our SPHERE observations. 
Azimuthal variations in surface density are regularly observed at longer (millimeter) wavelengths with ALMA, for example around \object{HD~142527} \citep{Perez2014} or \object{V1247~Ori} \citep{Kraus2017}.
These azimuthal asymmetries are hypothesized to originate from pressure bumps in the gas that trap large, millimeter-sized dust particles in the disk midplane. 
With SPHERE/IRDIS, however, we trace small, micron-sized dust particles at the disk surface which are much less affected by particle trapping in the disk \citep[see, e.g.,][]{Pinilla2016}. 
It is thus unlikely that we would observe an extreme asymmetry in scattered light. 
This effect can indeed be observed, for example for the \object{HD~142527} transition disk where the strong azimuthal asymmetry in large dust grains is not visible in scattered light (\citealt{Avenhaus2014}, \citealt{Casassus2015}).

This leaves us with the hypothesis that the northwestern side of the visible disk structure is possibly shadowed by an unresolved part of the disk at separations not probed by the SPHERE observations.
This can be the case if the inner part of the disk is misaligned with respect to the visible structures. 
For example, according to \citet{Price2018}, a (sub)stellar companion may cause this misalignment of an inner disk.
Such a misalignement can produce a variety of features from sharp, dark lanes, as observed in the disks around \object{HD~142527} \citep{Avenhaus2014}, \object{HD~100453} \citep{Benisty2017}, or \object{HD~135344B} \citep{Stolker2016}, to broader wedges, as reported for \object{PDS~66} \citep{Wolff2016} or \object{TW~Hya} \citep{Debes2017}. 
Recently, \citet{Benisty2018} showed scattered light images of the circumstellar disk around \object{HD~143006} in which, analogously to the current case, approximately half of the outer ring is shadowed by inner disk structures and is thus not detected in scattered light.
As presented in Sect.~\ref{subsec:sed_modeling}, the analysis of the object's SED strongly indicates the presence of a hot inner component of the disk, just as required for the proposed shadowing scenario.
However, the absence of narrow lanes implies that if present, the shadowing must be due to a very small misalignment.
To confirm this hypothesis, however, deeper data is required (e.g., a time series that could show the rotation of the shadowed regions around the star).

\subsection{Comparison with \object{HD~98363}}
\label{subsec:comparison_hd98636}

As studied by \citet{Chen2012} and \citet{Moor2017}, the primary star \object{HD~98363} hosts a gas-poor debris disk.
The detected disk around \object{Wray~15-788}, however, rather seems to be a gas-rich protoplanetary disk.

This brings up interesting questions about the evolution of the systems.
Assuming both formed at approximately the same time and with similar initial conditions, it is peculiar that the disk around \object{HD~98363} is already more evolved compared to the one around \object{Wray~15-788}.
As studied by \citet{Ribas2015}, there seems to be a trend of decreasing protoplanetary disk lifetimes with increasing mass of the star.
In their empirical study, however, they only compare the evolutionary stages of disks around stars above and below 2\,$M_\sun$.
Because both \object{Wray~15-788} and \object{HD~98363} fall into the latter category, their conclusions cannot directly be applied to our sample.
Another explanation for the different nature of the disks around the two stars might be the presence of multiple planetary companions around \object{Wray~15-788}.
These companions can act as traps for dust particles leading to a radial segregation of different sized dust particles as studied by \citet{Pinilla2015}.
To further explore possible scenarios, additional data on \object{Wray~15-788} is necessary.

\section{Conclusions}
\label{sec:conclusions}

For the first time, we resolved a transition disk around young K3IV star \object{Wray~15-788} in scattered light with both SPHERE/CI and SPHERE/DPI data.
SED analysis suggests that the star hosts a hot inner disk located interior to the IWA of the presented imaging data.
An excess at wavelengths longer than 10\,$\mu$m indicates additional disk material at larger separations from the star.
In agreement with this far-infrared SED, we identified an arc at a projected separation of $\sim$370\,mas and a potential inner ring at $\sim$170\,mas in the SPHERE data.
These two features are separated by a resolved region of significantly reduced flux.
From the outer arc, which is detected above 5$\sigma$ within a range of position angles of 60\degr\,\la\,$\phi$\,\la\,240\degr , we determined a disk inclination of $i$\,=\,21\degr\,$\pm$\,6\degr\ and a position angle of $\varphi$\,=\,76\degr$\,\pm\,$16\degr.
Correction for this inclination places the outer ring, the gap, and the inner substructures from the imaging data at approximate physical separations of 56\,au, 35\,au, and 28\,au, respectively.

Although we detected the disk at low inclination, large parts of the the outer ring remain hidden below the background noise.
This peculiar appearance may be caused by a shadow that is cast from unresolved inner substructures that are misaligned with respect to the outer material.
This scenario is in very good agreement with the SED of \object{Wray~15-788}, which shows clear evidence of an inner disk with an effective temperature of 969\,K.
The misalignment of this inner disk may be caused by an undetected substellar companion. 
From our 5$\sigma$ detection thresholds we derive an upper mass limit of 10\,$M_\text{jup}$ for a companion inside the detected gap.
At projected separations larger than 1\arcsec\ we can rule out companions more massive than 4\,$M_\text{jup}$; however, we cannot rule out the possibility that half of the disk is faint in the northwest and that our S/N is not high enough to detect it.

Furthermore, we found \object{Wray~15-788} to be companion to the A2V star \object{HD~98363}.
Therefore, \object{Wray~15-788} is actually \object{HD~98363~B} at a separation of $\sim$50\arcsec ($\approx$\,6900\,au) to the primary.
Even though both objects have the same age of $11^{+16}_{-7}$\,Myr, the primary hosts a debris disk where most of the primordial gas has already dissipated, whereas we are confident to detect a less evolved protoplanetary disk around \object{Wray~15-788}.
Possible undetected companions may be responsible for trapping the dust, leading to the different kind of disks within the binary system of \object{HD~98363} and \object{Wray~15-788}.

Further, deeper observations need to be conducted to better understand the disk's peculiar morphology and to find possible planetary-mass companions.
To confirm the detection of an inner ring and to constrain the inclination of the disk, a deeper, polarimetric observation is necessary.
Additional constraints to the disk's composition, the presence of gas, and the sizes of its dust grains can be set with submillimeter observations making use of the Atacama Large Millimeter/submillimeter Array (ALMA).

\begin{acknowledgements}
We thank C.~Dominik for his extremely valuable input on infrared excesses of circumstellar disks and the anonymous referee for providing useful feedback that helped to improve the quality of this article.
The research of AJB and FS leading to these results has received funding from the European Research Council under ERC Starting Grant agreement 678194 (FALCONER).

MB acknowledges funding from ANR of France under contract number ANR-16-CE31-0013 (Planet Forming disks).

Part of this research was carried out at the Jet Propulsion Laboratory, California Institute of Technology, under a contract with the National Aeronautics and Space Administration.

This publication makes use of VOSA, developed under the Spanish Virtual Observatory project supported from the Spanish MINECO through grant AyA2017-84089.

This research has made use of the SIMBAD database, operated at CDS, Strasbourg, France \citep{Wenger2000}.

To achieve the scientific results presented in this article we made use of the \emph{Python} programming language\footnote{Python Software Foundation, \url{https://www.python.org/}}, especially the \emph{SciPy} \citep{SciPy}, \emph{NumPy} \citep{numpy}, \emph{Matplotlib} \citep{Matplotlib}, \emph{scikit-image} \citep{scikit-image}, \emph{scikit-learn} \citep{scikit-learn}, \emph{photutils} \citep{photutils}, and \emph{astropy} \citep{astropy_1,astropy_2} packages.

\end{acknowledgements}

\bibliographystyle{aa} 
\bibliography{mybib} 

\begin{appendix}

\section{Reference star library}
\label{sec:ref_star_library}

To remove both stellar halo and instrumental artifacts in the data on \object{Wray~15-788} obtained with SPHERE in CI mode, we made use of an approach based on RDI in combination with PCA.
The stars used for our reference library are all young, K-type star members of LCC subgroup of Sco-Cen.
We list the names and corresponding observational parameters in Table \ref{tbl:ref_star_observations}.
\begin{table*}
\caption{Observations of reference stars carried out with SPHERE/IRDIS. All data were obtained in classical imaging mode.}
\label{tbl:ref_star_observations}
\def\arraystretch{1.2}
\setlength{\tabcolsep}{12pt}
\centering
\begin{tabular}{@{}lllllll@{}}
\hline\hline
Target & Observation date & Filter\tablefootmark{a} & NDIT$\times$DIT\tablefootmark{b} & $\langle\omega\rangle$\tablefootmark{c} & $\langle X\rangle$\tablefootmark{d} & $\langle\tau_0\rangle$\tablefootmark{e} \\ 
(2MASS ID) & (yyyy-mm-dd) & & (1$\times$s) & (\arcsec) & & (ms)\\
\hline
J11272881-3952572 & 2017-04-18 & \textit{H} & 4$\times$32 & 1.51 & 1.10 & 1.40\\
J11320835-5803199 & 2017-06-17 & \textit{H} & 4$\times$32 & 0.67 & 1.47 & 2.90\\
J11445217-6438548 & 2018-05-14 & \textit{H} & 4$\times$32 & 0.73 & 1.31 & 2.38\\
J11445217-6438548 & 2018-05-14 & \textit{K}$_\textit{s}$ & 4$\times$32 & 0.78 & 1.31 & 2.60\\
J12065276-5044463 & 2017-04-02 & \textit{H} & 3$\times$32 & 1.24 & 1.12 & 1.50\\
J12090225-5120410 & 2018-05-15 & \textit{H} & 4$\times$32 & 0.86 & 1.12 & 2.70\\
J12090225-5120410 & 2018-05-15 & \textit{K}$_\textit{s}$ & 4$\times$32 & 0.70 & 1.12 & 2.90\\
J12101065-4855476 & 2017-04-18 & \textit{H} & 4$\times$32 & 1.71 & 1.15 & 1.40\\
J12123577-5520273 & 2017-06-17 & \textit{H} & 4$\times$32 & 0.77 & 2.41 & 2.80\\
J12185802-5737191 & 2017-06-17 & \textit{H} & 2$\times$32 & 0.72 & 1.22 & 2.70\\
J12220430-4841248 & 2017-04-18 & \textit{H} & 3$\times$32 & 1.82 & 1.17 & 1.40\\
J12234012-5616325 & 2017-06-17 & \textit{H} & 4$\times$32 & 0.63 & 1.73 & 3.45\\
J12393796-5731406 & 2017-06-17 & \textit{H} & 4$\times$32 & 0.64 & 1.77 & 3.83\\
J12404664-5211046 & 2018-04-30 & \textit{H} & 4$\times$32 & 0.75 & 1.13 & 7.05\\
J12404664-5211046 & 2018-04-30 & \textit{K}$_\textit{s}$ & 4$\times$32 & 0.87 & 1.13 & 7.10\\
J12454884-5410583 & 2018-04-30 & \textit{H} & 4$\times$32 & 0.71 & 1.15 & 6.93\\
J12454884-5410583 & 2018-04-30 & \textit{K}$_\textit{s}$ & 4$\times$32 & 0.66 & 1.15 & 8.98\\
J12480778-4439167 & 2017-06-17 & \textit{H} & 4$\times$32 & 0.90 & 1.34 & 2.75\\
J13055087-5304181 & 2018-07-04 & \textit{H} & 4$\times$32 & 0.82 & 1.14 & 1.95\\
J13055087-5304181 & 2018-07-04 & \textit{K}$_\textit{s}$ & 4$\times$32 & 0.93 & 1.14 & 2.03\\
J13064012-5159386 & 2018-04-30 & \textit{H} & 4$\times$32 & 0.56 & 1.13 & 8.15\\
J13064012-5159386 & 2018-04-30 & \textit{K}$_\textit{s}$ & 4$\times$32 & 0.56 & 1.13 & 9.88\\
J13065439-4541313 & 2018-04-08 & \textit{H} & 4$\times$32 & 0.46 & 1.09 & 5.65\\
J13065439-4541313 & 2018-04-08 & \textit{K}$_\textit{s}$ & 4$\times$32 & 0.55 & 1.09 & 4.68\\
J13095880-4527388 & 2018-05-01 & \textit{H} & 4$\times$32 & 1.08 & 1.07 & 2.70\\
J13095880-4527388 & 2018-05-01 & \textit{K}$_\textit{s}$ & 4$\times$32 & 1.03 & 1.07 & 2.45\\
J13103245-4817036 & 2018-05-01 & \textit{H} & 4$\times$32 & 1.03 & 1.10 & 3.30\\
J13103245-4817036 & 2018-05-01 & \textit{K}$_\textit{s}$ & 4$\times$32 & 0.87 & 1.10 & 4.40\\
J13121764-5508258 & 2018-05-15 & \textit{H} & 4$\times$32 & 0.62 & 1.16 & 2.50\\
J13121764-5508258 & 2018-05-15 & \textit{K}$_\textit{s}$ & 4$\times$32 & 0.62 & 1.16 & 3.00\\
J13174687-4456534 & 2018-05-28 & \textit{H} & 4$\times$32 & 0.70 & 1.07 & 4.33\\
J13174687-4456534 & 2018-05-28 & \textit{K}$_\textit{s}$ & 4$\times$32 & 0.67 & 1.07 & 4.15\\
J13233587-4718467 & 2017-04-02 & \textit{H} & 4$\times$32 & 1.68 & 1.21 & 1.40\\
J13334410-6359345 & 2017-07-05 & \textit{H} & 4$\times$32 & 1.06 & 1.53 & 3.05\\
J13354082-4818124 & 2017-04-02 & \textit{H} & 4$\times$32 & 1.06 & 1.30 & 2.08\\
J13380596-4344564 & 2017-04-02 & \textit{H} & 4$\times$32 & 1.05 & 1.33 & 2.40\\
J13455599-5222255 & 2018-04-28 & \textit{H} & 4$\times$32 & 0.64 & 1.13 & 6.35\\
J13455599-5222255 & 2018-04-28 & \textit{K}$_\textit{s}$ & 4$\times$32 & 0.65 & 1.13 & 6.03\\
\hline
\end{tabular}
\tablefoot{
\tablefoottext{a}{A broadband filter in either \textit{H}- or \textit{K}$_\textit{s}$-band was applied.}
\tablefoottext{b}{NDIT describes the number of dithering positions and DIT is the detector integration time per dithering position.}
\tablefoottext{c}{$\langle\omega\rangle$ denotes the average seeing conditions during the observation.}
\tablefoottext{d}{$\langle X\rangle$ denotes the average airmass during the observation.}
\tablefoottext{e}{$\langle\tau_0\rangle$ denotes the average coherence time during the observation.}
}
\end{table*}
We observed 26 and 12 reference stars (PI: M.~A.~Kenworthy) in \textit{H} and \textit{K}$_\textit{s}$ band, respectively.
The same observational setup as for the science data on \object{Wray~15-788} was used.

\section{Other reduction strategies}
\label{sec:other_reductions}

In addition to the results presented in section \ref{sec:results}, we apply other data reduction strategies for both SPHERE/CI and SPHERE/DPI data.
In this way we can test the stability of the detected disk's appearance and morphology.

\subsection{CI data}
\label{subsec:other_reductions_ci}

Figure \ref{fig:other_reductions_ci} shows the individual analysis of the dataset on \object{Wray~15-788}.
\begin{figure*}
\resizebox{\hsize}{!}{\includegraphics{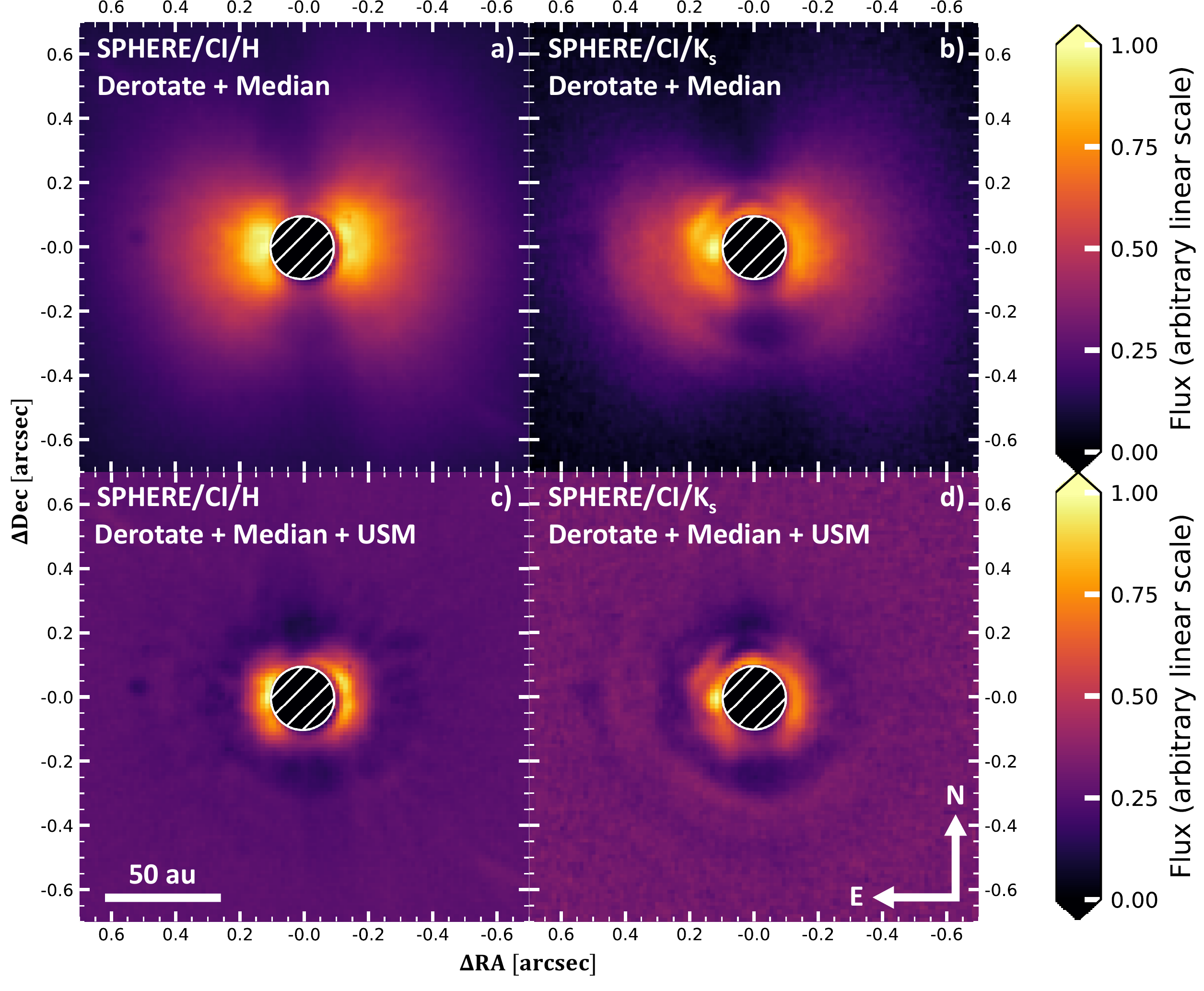}}
\caption{
Other reductions of SPHERE classical imaging data on \object{WRAY~15-788}.
The images were normalized to the maximum in each frame and arbitrarily, linearly scaled.
No subtraction of the stellar point-spread function (PSF) was performed.
For all images north is up and east is left.
Images \textbf{a} and \textbf{b} show the de-rotated and median combined image stack of science frames in \textit{H} and \textit{K}$_\textit{s}$ band, respectively. 
Frames \textbf{c} and \textbf{d} represent the same image as above, but an additional unsharp mask (USM) is applied (Gaussian kernel with full width at half maximum equal to PSF size).
}
\label{fig:other_reductions_ci}
\end{figure*}
We did not subtract any PSF model, but only de-rotated the images to have north pointing up and east towards the left.
The median combined image of the four exposures is presented.
In frames \textbf{a} and \textbf{b} we show this result for $H$ and $K_s$ band, respectively.
There is no obvious detection of the disk in $H$ band; instead, the outer ring and the gap are marginally visible in the $K_s$-band result.
Furthermore, the brightness asymmetry from northwest towards the southeastern part can be marginally recovered as well.
To obtain the results presented in frames \textbf{c} and \textbf{d} of Fig.~\ref{fig:other_reductions_ci}, we also applied an unsharp mask to the results from the top panel of the figure.
For unsharp masking, we used a Gaussian kernel with a FWHM of the instrumental PSF size of 50\,mas and 55\,mas in $H$ band and in $K_s$ band, respectively.
Due to this high-pass filtering we are able to detect some structures of the outer ring in both filters.
Also, the gap of the disk is highlighted. 

\subsection{DPI data}
\label{subsec:other_reductions_dpi}

In addition to the linear Stokes parameters $Q$ and $U$, we derived their azimuthal analogs $Q_\phi$ and $U_\phi$ according to \citet{Schmid2006} as
\begin{align}
\label{eqn:Q_phi}
Q_\phi &= Q\cos\left(2\phi\right) + U\sin\left(2\phi\right),\\  
\label{eqn:U_phi}
U_\phi &= Q\cos\left(2\phi\right) - U\sin\left(2\phi\right),  
\end{align}
where $\phi$ denotes the position angle as defined before.
By construction, $Q_\phi>0$ refers to a polarization direction azimuthally oriented around the star.
This is what we expect from stellar flux being recorded on the detector after a single scattering event at the dust grains of the disk.
A negative value of $Q_\phi$, however, represents a polarization vector radially aligned to the star.
The $U_\phi$ image can be used as a measure of an upper limit on the noise inside the $Q_\phi$ frame.

We present the azimuthal Stokes vectors in Fig. \ref{fig:other_reductions_pol}.
\begin{figure*}
\resizebox{\hsize}{!}{\includegraphics{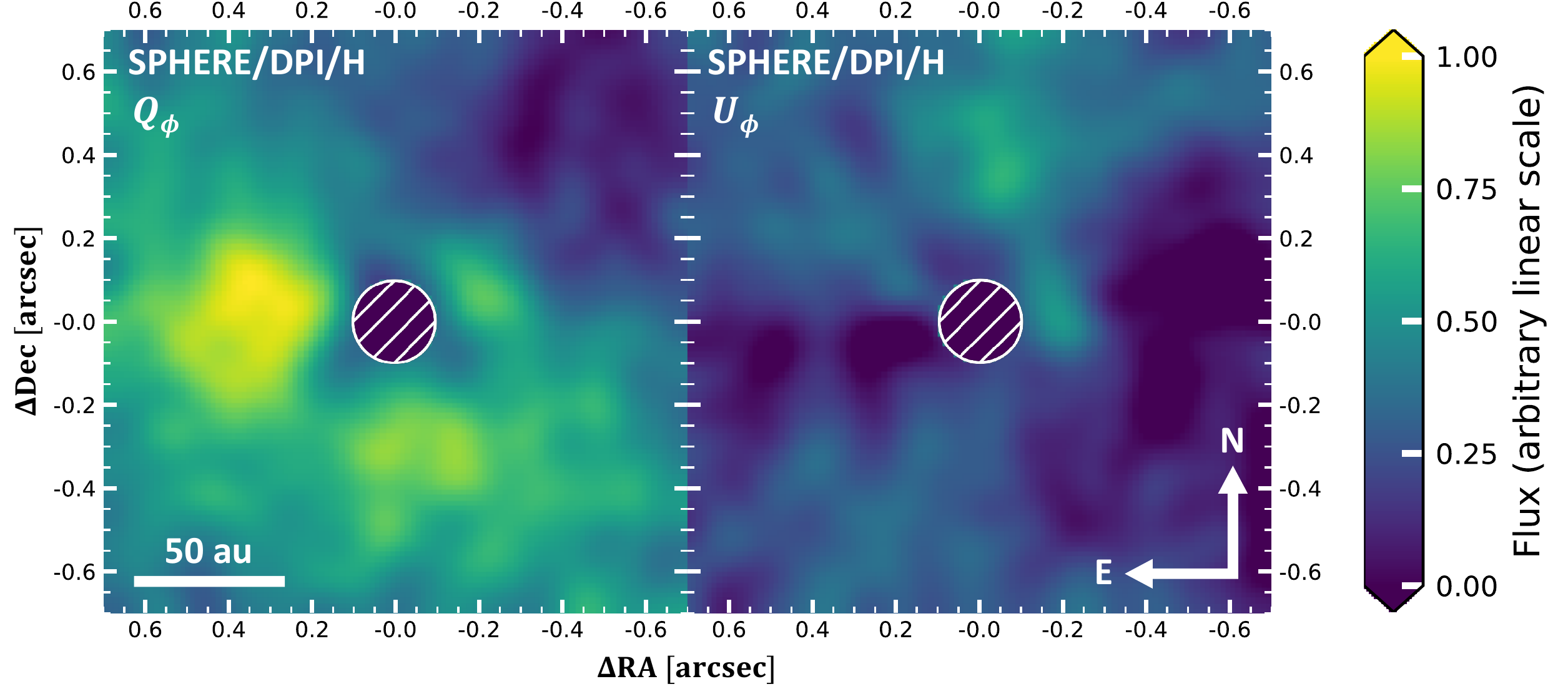}}
\caption{
Other reductions of SPHERE dual-polarimetric imaging data on \object{WRAY~15-788}.
We present the azimuthal Stokes parameters $Q_\phi$ and $U_\phi$ in the left and right panel, respectively.
Both images are scaled with $r^2$ and normalized with respect to the minimum and maximum flux in the $Q_\phi$ frame. 
In both frames north is up and east is left.
}
\label{fig:other_reductions_pol}
\end{figure*}
The images are smoothed with a Gaussian kernel having a FWHM of the PSF size and are scaled with the squared radial distance to the image center.
Both frames are normalized to the minimum and maximum value of the $Q_\phi$ image.

In the $Q_\phi$ image, we detect a clear indication of azimuthally polarized flux southeast of the star.
This agrees very well with our other observations from section \ref{sec:results}.
However, due to the non-optimal observing conditions the noise level is rather high, as seen in the $U_\phi$ frame.
Therefore, the data is not suited to study the morphology of the disk or to quantitatively compare it to the CI results presented in the top panel of Fig.~\ref{fig:WRAY15788_result}.

\end{appendix}

\end{document}